\documentclass{article}
\usepackage{amssymb}
\usepackage{epsfig}

\begin{document}
\hfill ITFA-2004-29\\
\begin {center}

 {\Large\bf To be or not to be?}\\{\large\bf Magnetic monopoles in
 non-abelian gauge theories}\footnote{Contribution to ``Fifty Years of
 Yang-Mills Theory'', edited by G.'t Hooft}\\[1cm]

 {\large F. Alexander Bais\footnote{E-mail: bais@science.uva.nl}\\ 
Institute for Theoretical Physics\\ 
University of Amsterdam}\\[5mm]

\end{center}


\begin{abstract}
Magnetic monopoles form an inspiring chapter of theoretical
physics, covering a variety of surprising subjects. We review
their role in non-abelian gauge theories. An expos\'e of exquisite
physics derived from a hypothetical particle species, because the
fact remains that in spite of ever more tempting arguments from
theory, monopoles have never reared their head in experiment. For
many relevant particulars, references to the original literature
are provided.

\end{abstract}

\section{Introduction}
\begin{quote}
{\it "Under these circumstances one would be surprised if Nature
had made no use of it"\\\mbox{}\hfill P.A.M. Dirac (1931)}
\end{quote}
\noindent The homogeneous Maxwell equations
\[ \nabla \cdot \mathbf{B}=0 \;\;\;,\;\;\;\;\nabla \times
\mathbf{E} + \frac{\partial \mathbf{B}}{\partial t}=0 \;\;,\] have
no source terms, reflecting the plain fact that isolated magnetic
poles or magnetic currents have never been observed in nature. The
experimental limit for observing heavy (non-relativistic)
monopoles in cosmic rays is presently well below $10^{-15} cm^{-2}
sr^{-1} sec^{-1}$, whereas on the other hand accelerator searches
have not produced any candidate up to masses of well over $500
\;GeV/c^2$. From a theoretical point of view this is surprising
because the absence of monopoles introduces an asymmetry in the
equations for which there does not appear to be any intrinsic
reason. On the contrary, when Dirac in his seminal 1931 paper
\cite{Dirac:1931}, introduced the magnetic monopole and studied
the consequences of its existence in the context of ordinary
quantum mechanics he did the striking discovery that the product
of electric and magnetic charges had to be quantized (with
$\hbar=c=1$),
\[ eg= 2\pi n \;,\] implying that the existence of a single monopole
would explain the observed quantization of all electric charges. In his
paper he introduces the magnetic "Dirac" potential,
\begin{eqnarray}\label{diracpotential}
  e\mathbf{A}^D (\mathbf{r})&=& \frac{eg}{4\pi}\;
  \mathbf{a}_{\hat{n}}(\mathbf{r})\nonumber \\
\mathbf{a}_{\hat{n}}(\mathbf{r}) &=& \frac{\mathbf{\hat{r}}\times
\mathbf{\hat{n}}}{r(1-\mathbf{\hat{r}}\cdot \mathbf{\hat{n}})}
\end{eqnarray}
which has besides the obvious singularity at the origin also a
\textit{string} singularity extending from the origin out along the
$\mathbf{\hat{n}}$ direction. It is the requirement that physical
charges should not be able to detect the string that enforces the
quantization condition. The quantization and conservation of magnetic
charge in electrodynamics has a topological origin and is related to
the existence of nontrivial circle bundles over the two-sphere
\begin{eqnarray}\label{u1bundle}
U(1)\hookrightarrow &B&  \nonumber \\
&\downarrow& \\
&S^2& \nonumber
\end{eqnarray}
These bundles are classified by the homotopy classes of of
mappings of a circle in the gauge group $U(1)$. It follows also
directly from the argument given by Dirac that the different
allowed magnetic charges are indeed in one to one correspondence
with these classes.  The simplest nontrivial example is the case
$B=S^3$, corresponding to the so-called Hopf fibration introduced
by the Hopf\cite{Hopf:1931} in 1931, the same year in which Dirac
wrote his monopole paper. This correspondence explains the
topological nature of magnetic charge and its conservation; the
addition or composition of charges is given by the homotopy group
$\pi_1(U(1))\simeq \pi_1(S^1)=\mathbb{Z}$. This topological
argument may be extended to classify the Dirac-type monopoles in
theories with non-abelian groups. For example in a pure gauge
theory (with only adjoint fields) the group is the simply
connected covering group divided by its center, e.g. $G\simeq
SU(N)/\mathbb{Z}_N$ and consequently $\pi_1(G) = \mathbb{Z_N}$,
implying that Dirac-type monopole charges are only conserved
modulo $N$. In spite of its attractive simplicity and elegance,
the Dirac analysis is vulnerable because monoples have to be
introduced by hand. A hypothetical particle which remains elusive
to this very day.

In the theoretical arena however, monopoles thrived for the more then
70 years that followed. The subject certainly went through various ups
and downs but it is fair to say that in spite of  the dramatic lack of
experimental evidence, the theoretical case has gained strength to the
point that monopoles appear to be unavoidable in any theory that wants
to truly unify electromagnetism with the other fundamental forces.
Whether it is through the road of Grand Unification or through the
compactification of spatial dimensions \'{a} la Kaluza and Klein and/or
through String Theory, monopoles appear to be the price one has to pay.

A magnificent impetus to the subject was the remarkable discovery
of 't Hooft and Polyakov\cite{'tHooft:1974qc,Polyakov:1974ek} that
in non-abelian gauge theories spontaneously broken down to some
$U(1)$ subgroup, magnetic monopoles appear as solitons, i.e. as
regular, finite energy solutions to the field equations. Monopoles
reappeared as natural and unavoidable inhabitants of the
non-abelian landscape; a package deal. And again the conservation
of magnetic charge arose as a consequence of the topology of the
solution space and not because of some symmetry argument.

After introducing some monopole essentials we discuss a variety of
topics that make monopole physics so fascinating, varying from
charge quantization issues and monopole stability, to making
fermions out of bosons. From theta angle physics to the catalysis
of baryon decay by Grand Unified Monopoles. From quantum moduli to
the links with a myriad of integrable systems...

Quantum Chromo Dynamics is a different domain where monopoles have
been used extensively in particular to explain the confinement
phenomenon. Here the challenge is to show that the vacuum consist
of monopoles that condensed to form a magnetically superconducting
ground state which confines quarks, as was suggested by Mandelstam
and 't Hooft. This is a statement that monopoles are literally
everywhere! How paradoxical nature can get? We will return to this
point when discussing the work of Seiberg and Witten on N=2
supersymmetric Yang-Mills Theory.
\section{Monopoles in Yang-Mills-Higgs Theories}
We consider spontaneously broken gauge theory - or the
Yang-Mills-Higgs system, with a gauge group $G$ and a Higgs field
$\Phi$ typically in the adjoint representation of the group. We
write the Lagrangian \footnote{Traces over squares of generators
are implied but not explicitly indicated if obvious. Spatial
vector quantities are printed in boldface}
\begin{equation}\label{action}
  \mathcal{L} = -\frac{1}{4}F_{\mu\nu}^{2}+(D_\mu\Phi)^{2}-\lambda
  V(\Phi)\; ,
\end{equation}
where
\begin{equation}\label{covariant}
D_{\mu}\Phi=\partial_{\mu}+e[A_{\mu},\Phi] \; .
\end{equation}
The symmetry is broken to a subgroup $H\subset G$ by a vacuum
expectation value $<\Phi>=\Phi_{0}$. For the solutions one imposes
the asymptotic condition at spatial infinity that each of the
terms vanishes sufficiently fast to have an integrable energy
density. This implies that $\Phi(\infty)\approx \Phi_{0}$ but also
that $D_{\mu}\Phi(\infty)\approx 0 $. We may also conclude that
\begin{equation}\label{longrange}
  [D_{\mu},D_{\nu}]\Phi\approx 0 \;,
\end{equation}
which implies that the only allowed long range (Coulombic type)
components for the gauge field $ F_{\mu\nu} \propto
[D_{\mu},D_{\nu}]$ have to lie in the unbroken part $H$.


\subsection{The BPS limit}

In the Lagrangian (\ref{action}) one may set the parameter $\lambda =
0$, which corresponds to the so-called
Bogomol'nyi-Prasad-Sommerfeld (BPS) limit. In this limit a lower
bound on the energy for a static, purely magnetic solution can be
derived\cite{Bogomolny:1976de}, the expression for the energy can
be casted in the form
\begin{equation}\label{bound}
  E = \frac{1}{2}\int  [{\bf B} \mp {\bf D}\Phi]^2 d^3x \pm \int
  {\bf B}\cdot {\bf D} \Phi d^3x \; .
\end{equation}
The last term can after a partial integration and using the
Bianchi identity be rewritten  ${\bf D}{\bf B}=0$ as,

\begin{equation}\label{mass}
  \pm\int {\bf B}\cdot {\bf D} \Phi \;\;d^3x = |g\Phi_0| =
  (\frac{eg}{4\pi})(\frac{4\pi}{e^2})M_w \; ,
\end{equation}
where $M_w = |e\Phi_0|$ is the mass of the charged vector
particles. This calculation shows that the mass of the monopole is
large: typically the mass scale in the theory divided by the fine
structure constant. The lower bound in (\ref{bound}) is saturated
if the fields satisfy the first order Bogomol'nyi equations
\begin{equation}\label{bogomolnyi}
{\bf B} = \pm{\bf D}\Phi
\end{equation}
This system of non-linear partial differential equations  has been
studied extensively and has become a most exquisite and rich
laboratory for mathematical physics, in particular with respect to
the study of higher dimensional integrable systems. We return to
this topic towards the end of the Chapter.

\subsection{The 't Hooft-Polyakov Monopole}

In 1974 't Hooft and Polyakov wrote their famous papers on the
existence of a regular magnetic monopole solution in the
Georgi-Glashow model with $G=SO(3)$ broken down to $U(1)$ by a
Higgs field in the triplet representation. Their argument implied
the existence of such regular monopoles in \textit{any} unified
gauge theory where the $U(1)$ of electromagnetism would be a
subgroup of some simple non-abelian group. The crucial difference
with the original Dirac proposal was that, because these monopoles
appear as regular, soliton-like solutions to the classical field
equations, they are unavoidable and cannot be left out. It is
amusing to note that it was originally thought that unifying
electromagnetism in a larger compact non-abelian group would
explain the quantization of electric charge \textit{without} the
need for magnetic monopoles.
\begin{figure}[!htb]
\begin{center}
\makebox[7.0cm]{\psfig{figure=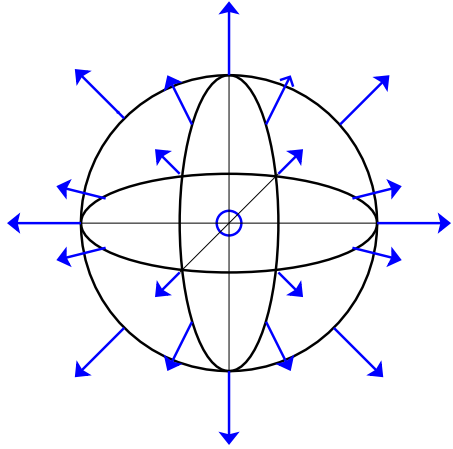,width=6.cm,angle=0}}
\caption[Hedgehog]{\footnotesize The topology of the isovector
Higgs field for the 't Hooft-Polyakov monopole configuration,
where the orientation in isospace is aligned with the position
vector in real space.} \label{hedgehog.eps}
\end{center}
\end{figure}
This basically because the charge
generator is identified with a \textit{compact} $U(1)$ generator
of the unified group. With the discovery of the regular monopoles
the two ways to obtain charge quantization  boil down to one and
the same argument. 't Hooft and Polyakov wrote down a spherically
symmetric \textit{ansatz} with respect to the mixed angular
momentum generator $\textbf{J=L+T}$, where $\mathbf{L} = -i \,
\mathbf{r} \times \nabla$ is the ordinary spatial part of angular
momentum and $\textbf{T}$ stands for the generators of the gauge
group. The symmetric \textit{ansatz} obeys the following natural
conditions:
\begin{eqnarray*}\label{spherical}
[J_i,\Phi] &=& 0 \\[0mm] [J_i, A_j] &=& i \varepsilon_{ijk} A_k
\end{eqnarray*}
and takes the form
\begin{eqnarray}\label{ansatz}
  \Phi &=& \frac{h(x)}{x} (\hat{\textbf{r}}\cdot \textbf{T})\nonumber \\
  \textbf{A} &=& \frac{1-k(x)}{x} (\hat{\textbf{r}}\times
  \textbf{T})
\end{eqnarray}
where we used the dimensionless variable $x \equiv efr$ with
$f\equiv |\Phi_0|$. Prasad and Sommerfield found an exact solution
\cite{Prasad:1975kr} in the BPS limit named after them.
\begin{equation}\label{prasad}
  h(x) = 1 - x\coth x \;\;,\;\;k(x)=\frac{x}{\sinh x} \;.
\end{equation}
It was later pointed out by Bogomol'nyi that these expressions are
a solution to a beautiful and much simpler system of the first
order equations (\ref{bogomolnyi}). Calculating the magnetic
charge by integrating the unbroken radial magnetic field component
at infinity, the intriguing result $g=4\pi/e$ came out - twice the
Dirac value. On the other hand, realizing that the monopole
solution could also be written down in a theory with fields in
doublet representations makes it less surprising, because these
fields carry $q=\pm e/2$. Dirac's veto is respected by all charges
in the theory.


\section{Charge quantization in non-abelian theories}

We remarked before that long range fields are allowed only in the
unbroken part of the gauge group. Simple time independent
solutions that for large $r$ have a purely magnetic monopole field
are
\begin{equation}\label{potential}
  eA = \frac{eg}{4\pi}(1+\cos\theta)d\varphi + O(\frac{1}{r^{2}})
\end{equation}
where the quantity $eg/4\pi$ is a constant element in the Lie
algebra of $H$, i.e. just a Dirac type potential for any component
in the unbroken group.

The extension of the  Dirac argument to the general non-abelian
case is rather straightforward. As the electrically charged fields
that may appear in the theory form representations of G one has to
impose that these are single valued if acted upon by the group
element $\exp eg$. This quantization condition can be solved in
terms of the simple roots $\vec{\gamma}_{i}$ (with $i=1,...,r$ and
$r$ is rank of $H$) of the root system of $H$.
\begin{figure}[!htb]
\begin{center}
\makebox[7.0cm]{\psfig{figure=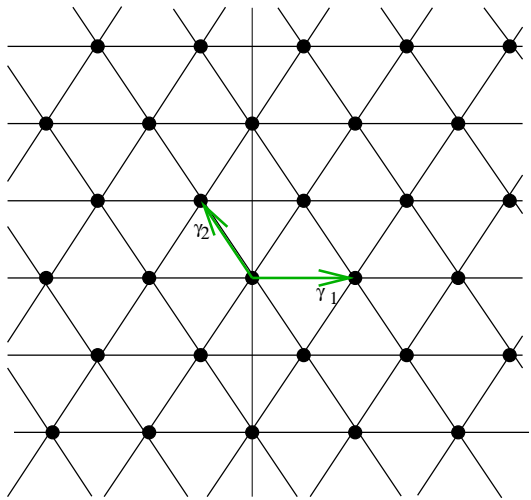,width=7.cm,angle=0}}
\caption[su3lattice]{\footnotesize The Lattice of
  allowed magnetic charges for the group $SU(3)$ spanned by the
  (inverse) simple roots.} \label{su3lattice0.eps}
\end{center}
\end{figure}
From the simple roots one constructs a convenient basis $\{C_i\}$
for the commuting generators of the Cartan subalgebra of $H$ with
the property that each basis element has per definition half
integral eigenvalues when acting on a basis vector of any
representation, this is achieved by defining,
\[ C_{i}\equiv
\frac{\vec{\gamma}_{i}}{|\vec{\gamma}_{i}|^2}\cdot\vec{C}\] where
$\{\vec{C}\}$ is the familiar Weyl basis for the Cartan
subalgebra. The solution to the quantization condition takes the
simple form:
\begin{equation}\label{quantisation}
  \frac{eg}{4\pi}=\sum^{r}n_i C_i
\end{equation}
This solution can be represented as a $r$-dimenional lattice dual
to the weight lattice of the group \cite{Englert:1976ng}. For the
simple example of $SO(3)$ the rank is r=1 and one gets indeed the
charges which are $eg=4\pi n$. For the group $SU(3)$ the
corresponding magnetic charge lattice is given in
Fig.\ref{su3lattice0.eps}. In general the dual weight lattice may
be thought of as the root lattice of a dual group and this
observation led Goddard, Olive and Nuyts to suggest the
existence of such a hidden dual symmetry
group\cite{Goddard:1977qe}. We note that the existence of
nonsingular solutions or even the asymptotic stability of the
allowed charges does of course not follow from the quantization
condition.


\subsection{Topological Charges}

Magnetic charges are conserved for topological reasons which
implies that some suitably defined lowest allowed charges should
be stable. It was recognized early on that the magnetic charge in
non-abelian gauge theories could be related to some topological
invariant defined by the asymptotic behavior of the Higgs field.
The asymptotic conditions allow us to consider the Higgs field as
a mapping from a closed surface at spatial infinity $\partial M$
into the gauge orbit of the particular vacuum solution $\Phi_0$,
this orbit or vacuum manifold is homeomorphic to the coset space
$G/H$.Well known are Coleman's Erice
lectures\cite{Coleman:1975qj} on the subject, which concerning
this particular topic were based on earlier papers a number of
Russian
authors\cite{Schwarz:1975to,Schwarz:1976ab,Monastyrsky:1974cj}.
So from
\begin{equation}\label{phimap}
  \Phi(\infty) : \partial M \rightarrow G/H
\end{equation}
it follows that these maps fall into inequivalent classes which
form the second homotopy group of the coset space denoted as
$\pi_2(G/H)$. These homotopy groups specify the composition rules
for the topological charges in a quite general way, and it is
worth saying a few things about them. The topological structure
resides in the fiber bundle
\begin{eqnarray}\label{bundle}
H \hookrightarrow &G&  \nonumber \\
&\downarrow& \\
&G/H& \nonumber
\end{eqnarray}
One may determine the homotopy structure by exploiting the
following exact sequence of mappings of homotopy groups:
\begin{equation}\label{homotopy}
  \cdots\rightarrow \pi_2(G)
  \rightarrow \pi_2(G/H) \rightarrow \pi_1(H) \rightarrow \pi_1(G) \rightarrow
  \pi_1(G/H)
  \rightarrow \cdots
\end{equation}
The exactness of the sequence refers to the fact that the image of
a given homomorphism equals the kernel of the next one in the
sequence. A theorem of Poincar\'{e} states that for all semisimple
groups $\pi_2(G)=0$, in which case the exactness of the sequence
implies that
\begin{equation}\label{coleman}
  \pi_2(G/H) \simeq Ker[ \pi_1(H)\rightarrow \pi_1(G)]
\end{equation}
If $G$ is simply connected ($\pi_1(G)=0$) the topological
classification boils down to the fundamental group of the unbroken
gauge group $H$, very much in line with the generalized
Dirac-argument of the previous section. This may be understood as
follows, we take a simply connected gauge group and choose
$\Phi_0$ to break the symmetry to the maximal torus $H =
U(1)^{\otimes r} \subset G$, then one obtains that $\pi_1(H) =
\mathbb{Z}^{\otimes r}$. In this case all allowed charges on the
dual weight lattice are topologically conserved. There is a
perfect matching between topological sectors and points on the
lattice determined by the quantization condition. The situation
for the 't Hooft-Polyakov case is an example of this but there is
one subtlety, because the gauge group $SO(3)$  is not simply
connected - $\pi_1(SO(3))=\mathbb{Z}_2$ , only the even elements
of $\pi_1(H)$ are in the Kernel of the homomorphism of
Eqn.(\ref{coleman}). This explains that in spite of the fact that
the the residual group is just $U(1)$ the quantization condition
is twice the one given by Dirac.

Topologically stable monopoles do not occur in the electro-weak
sector of the standard model because in the breaking of $G=SU(2)
\times U(1)$ by a complex doublet one finds  that $G/H \simeq S^3$
and consequently $\pi_2(S^3)=0$.

The situation changes drastically if the gauge group is only
partially broken, then one may expect that there are fewer
topologically conserved components to the magnetic charge. With a
single adjoint Higgs field one may break for example to a group
$H=U(1)\otimes K$ where K is semi-simple and simply connected in
which case $\pi_1(H)= \mathbb{Z}$ and there is only a single
component of the magnetic charge that is topologically conserved.
This is the situation that arises in most Grand Unified Theories,
for example the $SU(5)$ theory broken down to $SU(3)\times SU(2)
\times U(1)$. Monopoles with unbroken non-abelian symmetries are
still rather poorly understood. The monopoles are to be related
with conjugacy classes and the dyonic states do not fill complete
representations of the unbroken group, but only representations of
the centralizer of the magnetic charge vector. There is an
obstruction to implement the full unbroken group in the presence
of a magnetic charge - sometimes referred to as the problem of
implementing global color\cite{Balanchandra:1983bn,Nelson:1984fn}.
Whereas for a single monopole the issue is quite well understood,
if one is to study the multi-monopole sectors and the fusion
properties of these monopoles the situation appears far from
trivial\cite{Bais:1997qy}.

A final nontrivial class of topological magnetic charges arises if the
gauge group gets broken to {\it non-simply-connected} non-abelian
subgroups in which case the magnetic charge is only additively
concerved modulo some integer\cite{Bais:1988fn}. With a Higgs field in
the 6-dimensional representation of $SU(3)$ one may break $SU(3)$ to
$SO(3)$ (defined by the embedding where the triplet of $SU(3)$ goes
into the vector of $SO(3)$), then one obviously gets
$\pi_1(SO(3))=\mathbb{Z}_2$. The resolution to the mismatch between
the sets of allowed and topologically conserved charges has to be that
certain components of the magnetic charge vectors on the dual weight
lattice are unstable. This will be discussed in the following
subsection.

Other homomorphisms in the exact homotopy sequence
(\ref{homotopy}) have also physical content, for example to
determine what happens when monopoles would cross a phase
boundary\cite{Bais:1981nn}. One can imagine monopoles which are
formed at an early stage of the universe and one would then like
to know their fate if the universe subsequently goes through
various phase transitions. For example if one has two broken
phases with $G\subset H_1 \subset H_2$, the question is what
happens to the monopoles that are allowed in the $H_1$ phase when
they cross a boundary to the $H_2$ phase; will they be confined,
converted or just decay in the vacuum? The exact sequence tells us
for example that
\begin{equation}\label{sequence}
Im[\pi_1(H_2)\rightarrow\pi_1(H_1)]=
Ker[\pi_1(H_1)\rightarrow\pi_1(H_1/H_2)]
\end{equation}
Given the fact that $\pi_1(H_1/H_2)$ labels the topological
magnetic flux tubes in the $H_2$ phase the above equation
determines exactly which $H_1$ monopoles will be confined.


\subsection{Charge instabilities, a mode analysis}

In general there is a discrepancy between the lattice of allowed
magnetic charges defined by(\ref{quantisation}) and the
topologically conserved (if not stable) subset. These should
somehow be related to each other. One way to get insight in this
question is to study the asymptotic stability of the long range
magnetic Coulomb field of a charge on the lattice. Brandt and Neri
studied the fluctuations and did indeed find certain unstable
modes on this singular background \cite{Brandt:1979kk}.
\begin{figure}[!htb]
\begin{center}
\makebox[3.7cm]{\psfig{figure=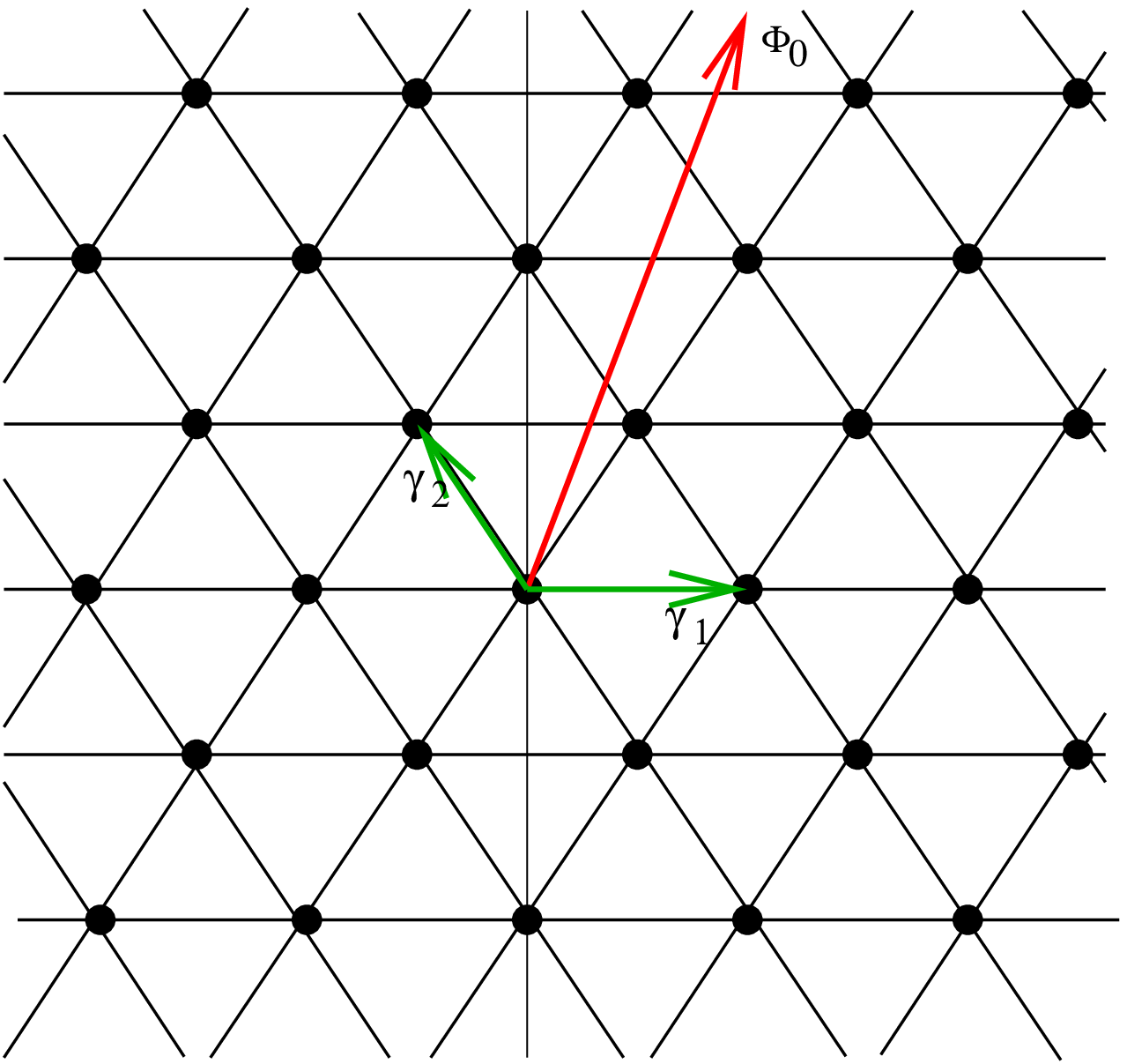,width=3.4cm,angle=0}}
\makebox[3.7cm]{\psfig{figure=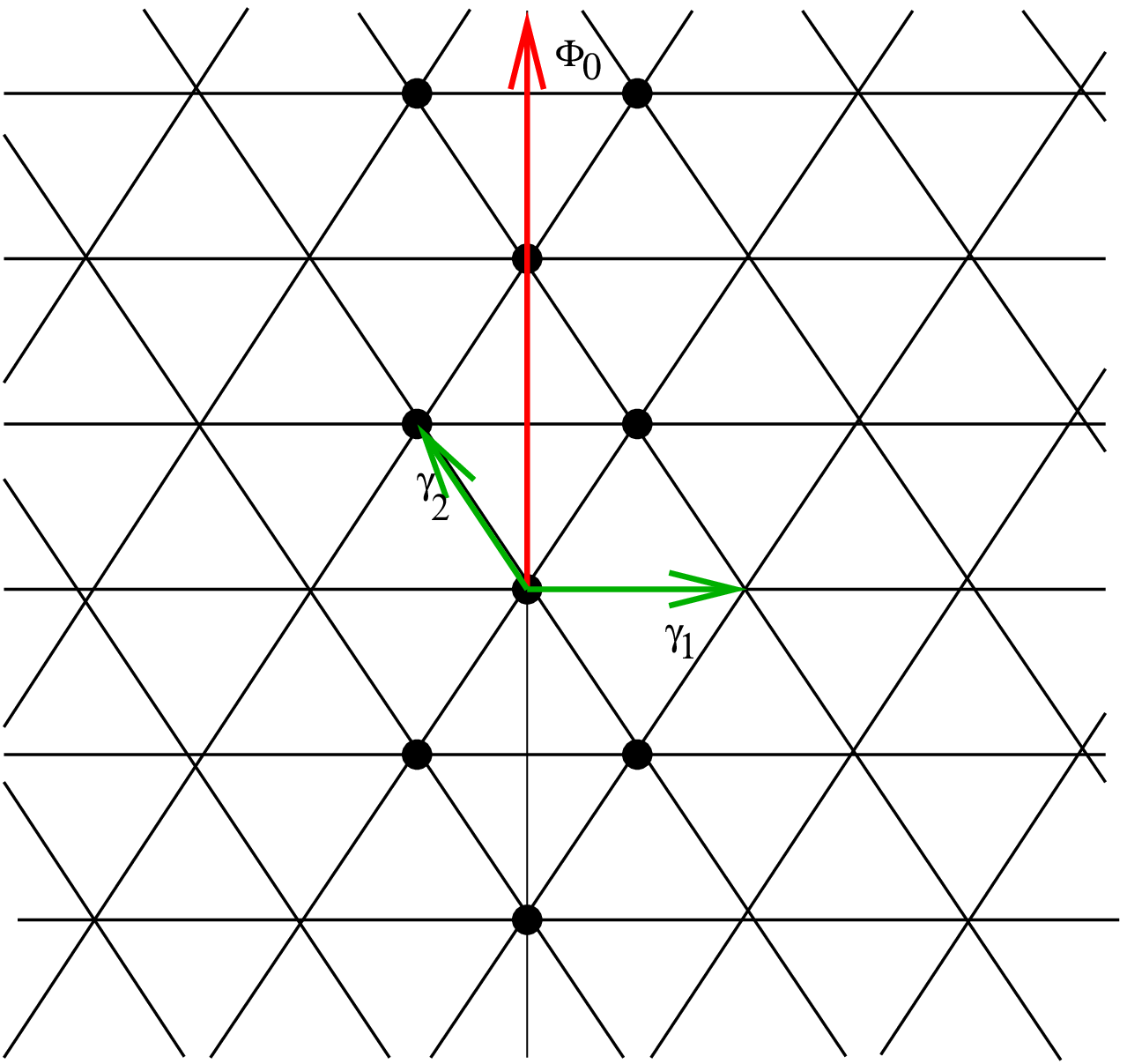,width=3.4cm,angle=0}}
\makebox[3.7cm]{\psfig{figure=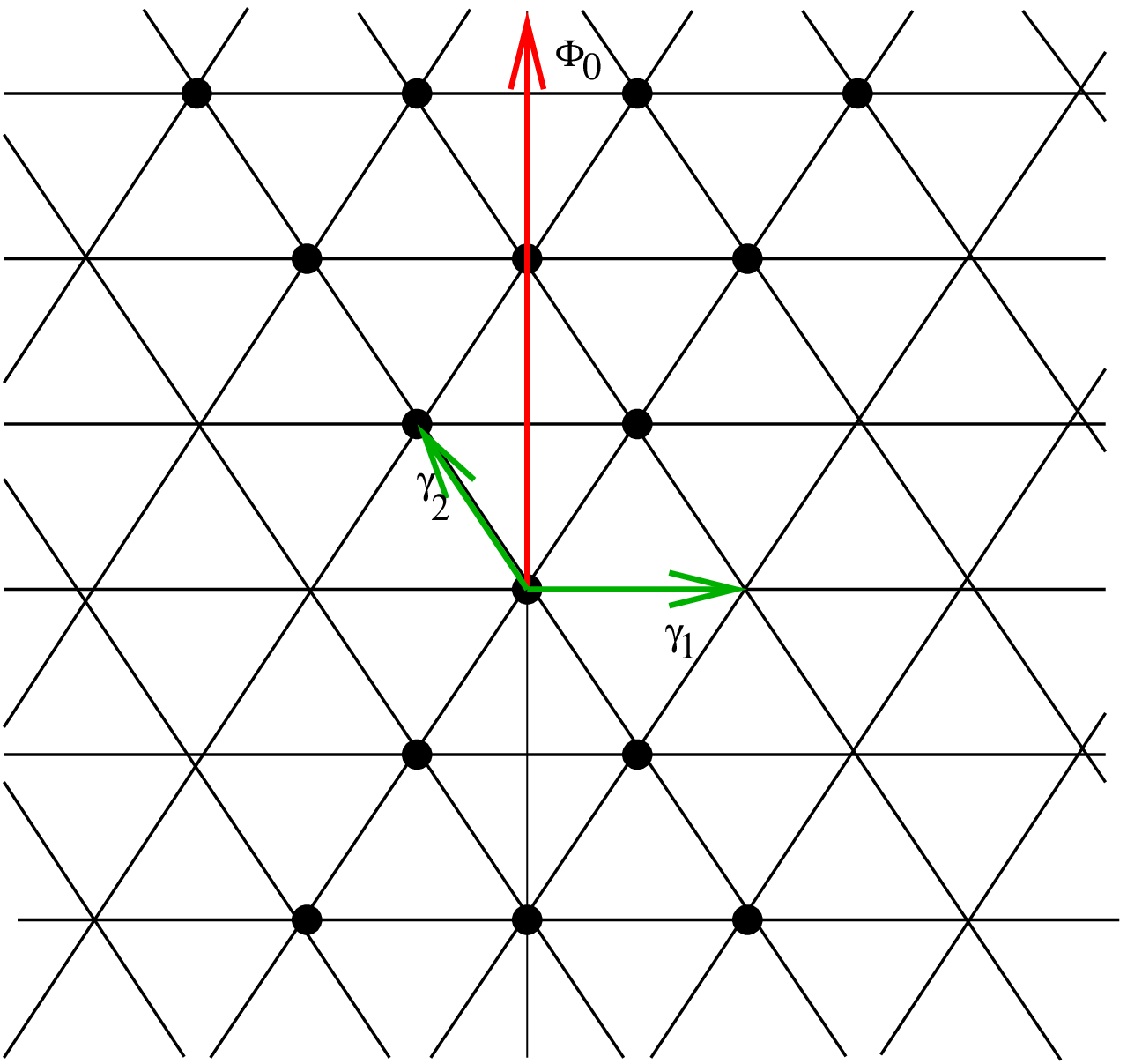,width=3.4cm,angle=0}}
 \makebox[3.7cm]{(a)} \makebox[3.7cm]{(b)}\makebox[3.7cm]{(c)}
\caption[somethingelse]{\footnotesize
The lattice of charges
allowed by the quantisation condition, spanned by the
  (inverse) simple roots,. In this figure also the direction of the Higgsfield $\Phi_0$ in the
  Cartan subalgebra is indicated. In (a) the
  stable charges  for an arbitrary
  non-degenerate orientation of the Higgs field are indicated by black dots. In that case the residual gauge
  group is $U(1)\times U(1)$ and all allowed charges
  are topologically conserved.  In (b) the Higgs field is degenerate
  and leaves the non-abelian group $U(2)$ unbroken. Now only one component of the
  magnetic charge is conserved, and in each topological sector only
  the smallest total charge is conserved. The points symmetric with
  respect to the Higgs field are gauge conjugates. In (c) we
  have indicated the same orientation of the Higgs field, but we
have  taken the Bogomoil'nyi limit and see that  more charges
  are stable then one would expect on purely topological grounds. The
  horizontal quantum number is sometimes referred to as the
  holomorphic charge.}
\label{su3lattice.eps}
\end{center}
\end{figure}
Let us expand the Lie algebra valued fields around their classical
backgrounds as:
\begin{eqnarray}\label{expansion}
  e{\bf A} &=& e{\bf A}_D + e{\bf a} \nonumber \\
  \Phi &=& \Phi_0 + \phi
\end{eqnarray}
where ${\bf A}_D$ is the Dirac monopole potential and work in the
background gauge ${\bf D}\cdot {\bf a} +[\Phi,\phi] = 0$.  It is
convenient to expand the commutator of the fluctuation field with
the monopole background and write,
\begin{equation}\label{commutator}
[\frac{eg}{4\pi},{\bf a}] = \sum_\alpha q(\alpha) {\bf a}^\alpha
T_\alpha
\end{equation}
from which it follows that $q(\alpha)=0$ if $T_\alpha$ is a
generator of the little group $S$ of the magnetic charge generator
$eg/4\pi$. Note that $S \cap H \neq 0$. In the BPS limit we should
replace $\Phi_0$ in the expansion (\ref{expansion}) by $\Phi_0 -
eg/4\pi r$ which yields extra $q$ dependent terms from commutators
with $\Phi$. The linearized fluctuation equations take the
following form:
\begin{eqnarray}
  ({\bf D}\cdot {\bf D}) {\bf a} + \frac{2q}{r^2}(\hat{{\bf r}}\times
  {\bf a})-\frac{2q}{r^2}\;\hat{{\bf r}}\;\phi -\frac{q^2}{r^2}\;{\bf a}
  - m^2 {\bf a} &=& - E^2 {\bf a} \nonumber \\
({\bf D}\cdot {\bf D})\phi-\frac{2q}{r^2}\;\hat{{\bf r}}\;{\bf
a}-\frac{q^2}{r^2}\;\phi - m^2\phi &=& -E^2 \phi
\end{eqnarray}
In these equations both fields carry a Lie algebra index $\alpha$
which is suppressed. The structure of this coupled system is now
as follows, the terms which are dependent on $q =q(\alpha)$ are
zero for the components which generate $S$, whereas the mass terms
$m^2 = m^2(\alpha)$ vanish for the components which generate $H$.
For components of the fluctuation fields outside $H\cup S$ there
will be no unstable modes because $E^2>0$. In the generic case,
not the BPS limit, all q dependent terms vanish except the second
term in the vector equation. This means that the equations
decouple in that case and one may verify that the scalar
perturbation has no unstable modes. The vector perturbations
however turn out to have one unstable mode for components inside
$S$ provided $\mid q \mid \geq 1$. In the generic case one arrives
therefore at an important conclusion which reconciles the notion
of topological charge and dynamical stability, namely, that
\textit{in each topological class only the smallest total magnetic
charge is stable}.
 The stability analysis in the BPS limit is
much more involved, indeed in that case there may be more than a
single stable monopole in a given topological sector. We have
illustrated various situations for $SU(3)$ in
Fig.\ref{su3lattice.eps}. In the first figure (a) we give the
complete lattice of allowed charges in $SU(3)$ all these are
topologically conserved if one breaks $SU(3)$ to $U(1)\times
U(1)$. The second figure (b) shows what happens if one breaks to
$U(2)$ (with a Higgs field along the $\lambda_8$ direction), then
the asymptotic analysis shows that only the black dots survive
while the others have become unstable. Indeed the minimal total
magnetic charge within each topological class survives, indicating
an instability in the horizontal direction. The last figure (c)
shows what happens in the Bogomol'nyi limit where the stability
analysis is affected by the massless scalar degrees of freedom,
now the stable monopoles fill out a Weyl chamber around the Higgs
direction.


\section{Cheshire charge and core instabilities}

It may happen that the topology of the vacuum manifold is more
complicated than the ones we just discussed. In particular it may
be such, that different types of defects can coexist. If the
residual symmetry group is non-abelian these defects may have
\textit{topological interactions}, interactions not mediated by
the exchange of particles, but interactions that are essentially
of a kinematical nature. Yet these interactions may in the end
lead to physical effects, like instabilities. The simplest
situation of this sort is encountered if one breaks the gauge
group to a non-abelian discrete group. The low energy description
of such a models is referred to as a \textit{discrete gauge
theory}. The defects are "magnetic" fluxes which however carry
non-abelian quantum numbers\cite{Bais:1980vd}. In a two
dimensional setting these would give rise to \textit{non-abelian
anyons}. Another class of models is obtained if the unbroken group
has several discrete components as well as some non-trivial first
homotopy group. In such situations there are monopoles as well as
topological fluxes and this may lead to remarkable physical
properties. The simplest example of this sort is \textit{Alice
electrodynamics} introduced by A.S. Schwartz in 1982
\cite{Schwarz:1982}. The original model  is just the
Yang-Mills-Higgs system (\ref{action}), with the Higgsfield $\Phi$
in the five dimensional, symmetric tensor representation of
$SU(2)$. The potential is given by:
\begin{equation}\label{alpotential}
V=-\frac{1}{2}\mu^2Tr\left(\Phi^2\right) -\frac{1}{3}\gamma
Tr\left(\Phi^3\right) + \frac{1}{4} \lambda
\left(Tr\left(\Phi^2\right)\right)^2\quad.
\end{equation}
and has three parameters. By a suitable choice of parameters the
Higgs field will acquire a vacuum expectation value, $\Phi_0$ of
the form $\Phi_0= diag(-f,-f,2f)$. It follows that the residual
gauge group $H=U(1)\ltimes\mathbb{Z}_2\sim O(2)$, so, in a certain
sense it is the most minimal non-abelian extension of ordinary
electrodynamics. The nontrivial $\mathbb{Z}_2$ transformation
reverses the direction of the electric and magnetic fields and the
sign of the charges.
\begin{equation}
X Q X^{-1}=-Q\quad,
\end{equation}
with $X$ the nontrivial element of $\mathbb{Z}_2$ and $Q$ the
generator of the $U(1)$.
\begin{figure}[!htb]
\begin{center}
\makebox[5.0cm]{\psfig{figure=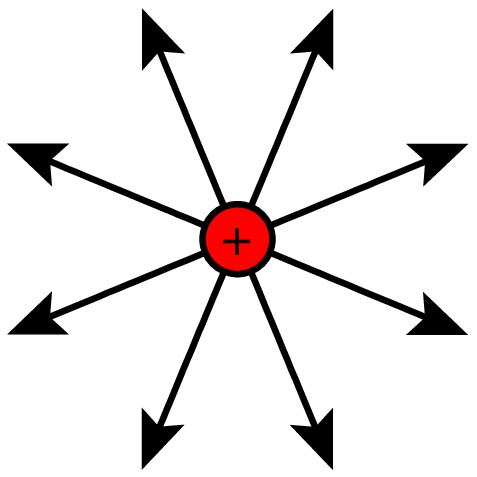,width=3cm,angle=0}}
\makebox[5.0cm]{\psfig{figure=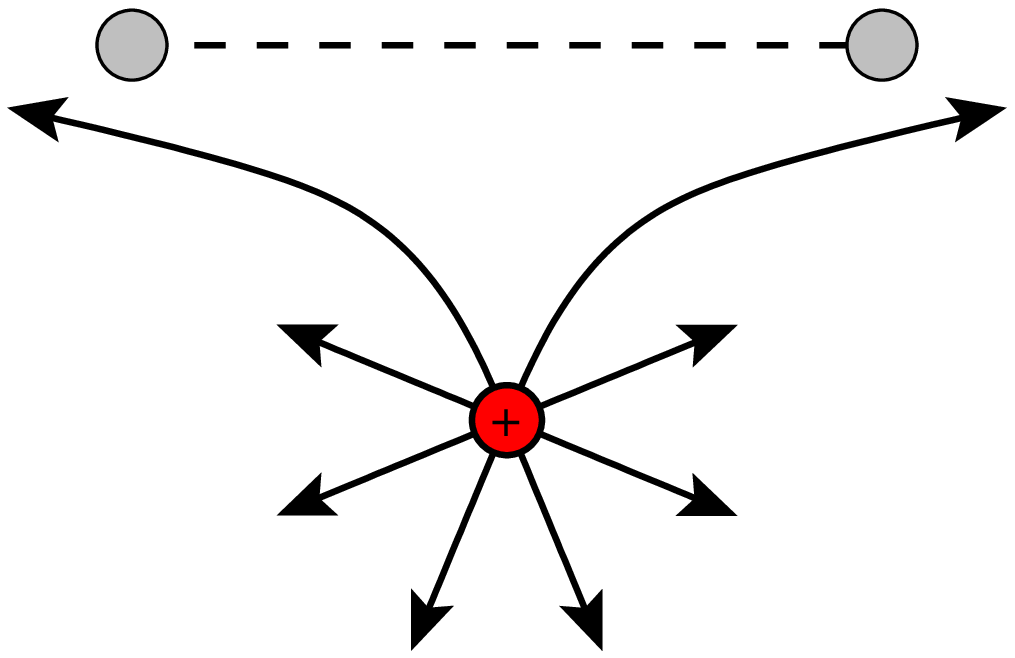,width=4cm,angle=0}}
\makebox[5.0cm]{(a)} \makebox[5.0cm]{(b)}
\makebox[5.0cm]{\psfig{figure=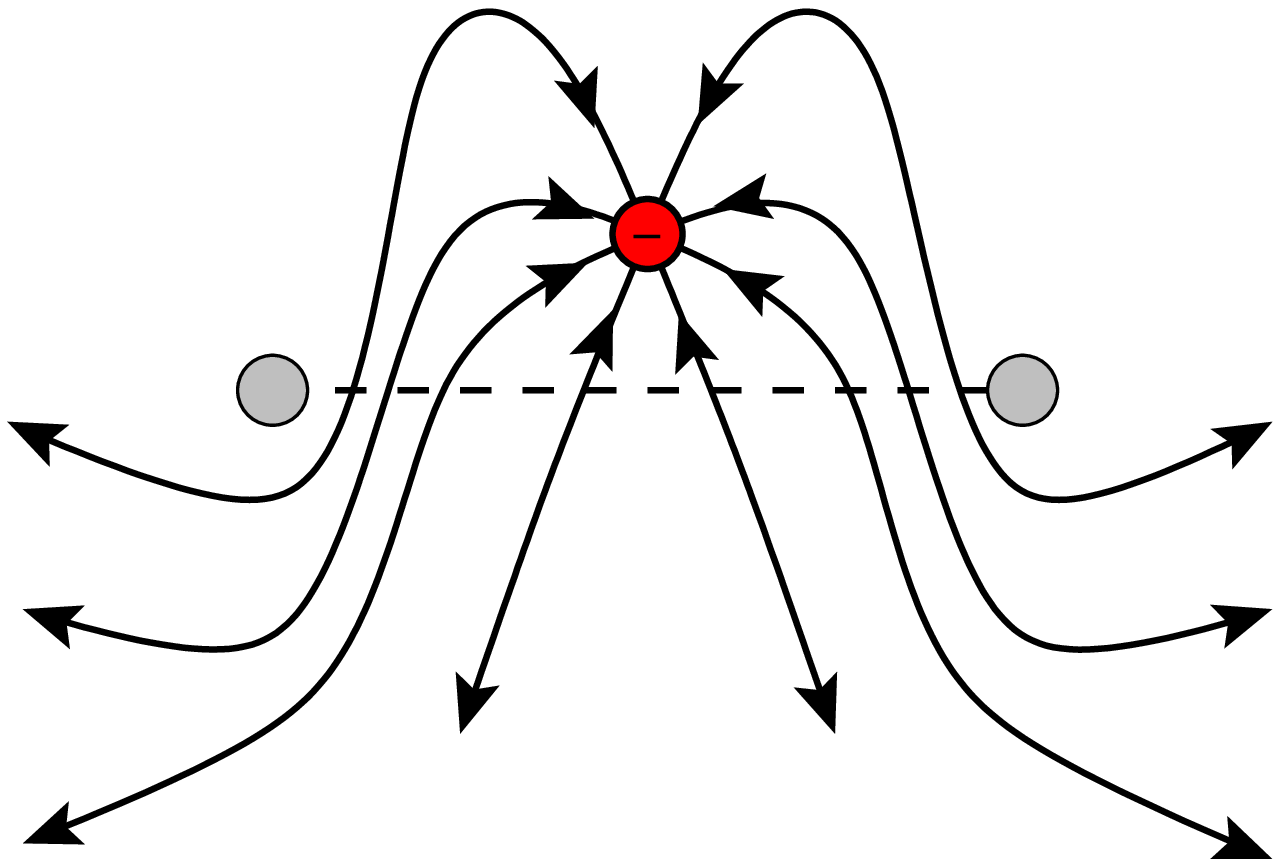,width=4cm,angle=0}}
\makebox[5.0cm]{\psfig{figure=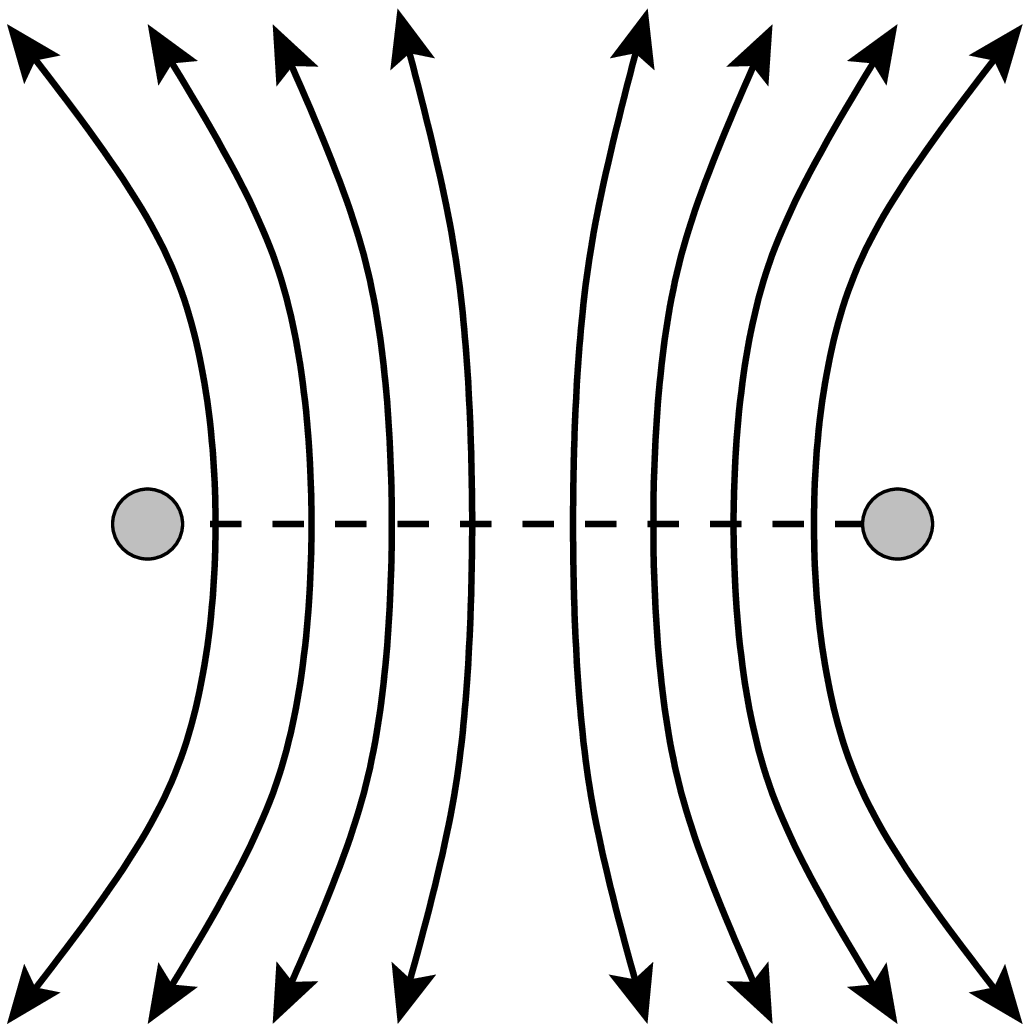,width=3cm,angle=0}}
\makebox[5.0cm]{(c)} \makebox[5.0cm]{(d)}
\caption[somethingelse]{\footnotesize This sequence of pictures
makes clear that the Cheshire phenomenon is generic in these
models  and does not depend on the particular symmetry of the
configuration. Using the fact that due to charge conservation
and/or quantization electric field lines cannot cross an Alice
flux one is lead to the notion of Cheshire charge.}
\label{chesmech.eps}
\end{center}
\end{figure}
The generator of $U(1)$ and the nontrivial element of the
$\mathbb{Z}_2$ do not commute with each other, in fact they
anti-commute. This means that the $\mathbb{Z}_2$ part of the gauge
group acts as a (local) charge conjugation on the $U(1)$ part of the
gauge group. Indeed, it is a version of electrodynamics in which charge
conjugation symmetry is gauged. From the structure of the (residual)
gauge group it is clear what the possible topological defects in this
theory are. As $\Pi_0(U(1)\ltimes\mathbb{Z}_2)=\mathbb{Z}_2$ there will
be a topological $\mathbb{Z}_2$ flux, denoted as \textit{Alice flux},
and furthermore as $\Pi_1(U(1)\ltimes\mathbb{Z}_2)=|\mathbb{Z}|$ there
are also magnetic monopoles in this theory (like in compact ED). The
element of the unbroken gauge group associated with the Alice flux
contains the nontrivial element of the $\mathbb{Z}_2$ part of the gauge
group, $X$. This means that if a charge is moved around an Alice flux
it gets charge conjugated. At first this might not be such a very
interesting observation as charge conjugation is part of the local
gauge symmetry of the model. However as mentioned before there is the
notion of a relative sign, which is path dependent in the presence of
Alice fluxes. This means that if one starts with two equal charges
(repulsion) and one moves one of the charges around an Alice flux one
ends up with two charges of the opposite sign (attraction), due to the
non-commutativity of $X$ and $Q$. This allows for the rather
interesting sequence of configurations depicted in
Fig.\ref{chesmech.eps}. where a charge is pulled through a ring of
Alice flux. Global charge conservation requires that it leaves behind a
(doubly) oppositely charged Alice ring, but charged in a peculiar
non-localizable way. The net charge in the region around the ring is
nonzero, yet a small test charge would be pulled through the ring
without encountering any source and then be pushed away on the other
side! The surface where the electric field lines change sign is a gauge
artefact, a fictitious Dirac sheet bounded by the Alice flux ring and
therefore does not carry real charge. Nevertheless, enclosing the whole
ring in a closed surface one measures a total net charge.  The possibility of this
type of non-localizable charge was first noted by
Schwarz\cite{Schwarz:1982}  and is referred to as \textit{Cheshire
charge}\cite{Alford:1991ab}, referring to the cat in \textit{Alice
in Wonderland} that disappears but leaves his grin behind. One may show
that - not surprisingly - this Cheshire property also holds for
magnetic charges\cite{Bucher:1992ab}.
\begin{figure}[!htb]
\begin{center}
\makebox[10.0cm]{\psfig{figure=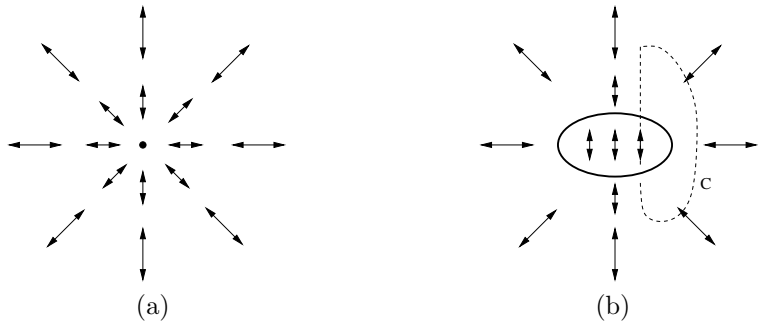,width=10cm,angle=0}}
\makebox[4.0cm]{(a)}\makebox[2cm]{} \makebox[4.0cm]{(b)}
\caption[Core deformation]{\footnotesize This figure illustrates
the possibility of a smooth deformation of a monopole core
topology of a point (a) to magnetically charged Alice ring
configuration (b).} \label{alice.eps}
\end{center}
\end{figure}
This is most directly illustrated by an allowed deformation of the
core topology of the monopole in this theory. This is shown
schematically in Fig.\ref{alice.eps}. We see that because of the
director  property (i.e. double arrowed nature) of the order
parameter field $\Phi_0$ it is possible to drill a hole through
the core maintaining continuity of the order
parameter\cite{Bais:1995zp}. The magnetic field lines stay
attached to the order parameter and spread out over the minimal
surface spanned by the ring, pretty much like ordinary magnetic
flux lines would spread when kept together by a super conducting
ring. Because of the presence of the additional parameter $\gamma$
in the potential (\ref{alpotential}), one can imagine that the
allowed deformation could lead to a dynamical core instability of
the 't Hooft-Polyakov monopole in this theory. It has been shown
that this is indeed the case; for a certain range of parameters
the magnetic Cheshire configuration has the lowest energy
\cite{Bais:2002ae}.


\section{Physics from moduli space}

The moduli space is basically the space of solutions $\mathcal{M}
= \{\mathbf{A,\Phi}\}$ modulo gauge transformations. This space of
physically inequivalent configurations has of course many
disconnected components $ \mathcal{M}_m$ labelled by the
topological charge $m$. This classically degenerate moduli space
can be described by collective coordinates, which upon
quantisation give a semiclassical spectrum of quantum states to
which the collection classical monopole solutions give rise. The
fact that monopoles can be located anywhere in real space leads to
translational zero modes. Modes associated with the compact
internal symmetry group will give rise to the "electric" gauge
charges that can be implemented in the various topological
sectors. The dimensionality of the moduli space in a given
topological sector can be computed by calculating the appropriate
index for the Dirac type operator in a background gauge. For the
$SU(2)$ Bogomol'nyi equations E. Weinberg\cite{Weinberg:1979ma}
constructed the operator and used an index theorem of
Callias\cite{Callias:1978kg} to calculate the dimension of
$\mathcal{M}_m$, the dimension of the moduli space (i.e. the
number of $L^2$ normalizable, independent zeromodes) in magnetic
sector $m$ with the result
\begin{equation}\label{index}
  dim\; \mathcal{M}_m = 4\mid m\mid \;.
\end{equation}
Roughly speaking this dimensionality can indeed be interpreted as
the number of degrees of freedom of $|m|$ individual fundamental
monopoles - each with three translational and one charge degree of
freedom. The $m=1$ moduli space has four parameters and the
geometry is simply $\mathcal{M}_1 = \mathbb{R}^3 \times S^1$. The
corresponding modes were constructed explicitly by Mottola
\cite{Mottola:1978pz}.

In the two-monopole moduli space one may go one step further and
discuss the low energy dynamics of monopoles. The isometric
decomposition of this space is
\begin{equation}\label{2monopole}
  \mathcal{M}_2 = \mathbb{R}^3 \times \frac{S^1 \times
  \bar{\mathcal{M}^0_2}}{\mathbb{Z}_2}
\end{equation}
The space $\bar{\mathcal{M}^0_2}$ is the double cover of the
moduli space of centered 2- monopole configurations, a
4-dimensional manifold named after Atiyah and Hitchin, who
determined the metric from the infinitesimal field modes, i.e. the
innerproduct of tangent vectors of this space
\cite{Gibbons:1995yw}. The manifold $\bar{\mathcal{M}^0_2}$ is
hyper-K\"ahler, and an anti-selfdual Euclidean Einstein space with
vanishing scalar curvature. The metric has furthermore an SO(3)
isometry group, and therefore can be written in terms of a radial
coordinate $r$ and three left invariant one-forms $\sigma_i\; (
i=1,2,3)$ that satisfy the relation $d\sigma_i =
\epsilon_{ijk}\sigma_j\sigma_k$. The metric becomes:
\begin{equation}\label{atiyahhitchin}
  ds^2 = (\frac{M_m}{2})_2 \left[ f(r)^2 dr^2 + a(r)^2\sigma_1^2 +
  b(r)^2\sigma_2^2 +c(r)^2\sigma_3^2\right].
\end{equation}
and approaches for large $r$ the Euclidean Taub-Nut metric. In the
moduli space approximation the classical scattering of two widely
separated monopoles is described by the geodesic motion on this
space \cite{Manton:1982mp,Atiyah:1985dv,Gibbons:1986df}. One may
go one step further and construct a Hamiltonian from the canonical
Laplacian on the modulispace to discuss the semi-classical bound
states and scattering cross sections of monopoles.

In the case of supersymmetric extensions there may be modes
associated with supersymmetries as well, which lead to
interesting consequences related with duality. We return to these
matters later on.


\subsection{Topologically non-trivial Gauge transformations}

In general one has the infinite dimensional group $\mathcal{G}$ of
smooth, time independent gauge transformations associated with the
structure group $G$
\begin{equation}\label{gauge}
  \mathcal{G}\equiv \{g : \mathbb{R}^3\rightarrow G\}.
\end{equation}

If the group $G$ is broken to a residual gauge group $H$ then we rather
like to consider the group $\mathcal{G}_r$ of \textit{residual gauge
transformations} defined as those transformation which leave the
asymptotic Higgs field invariant
\begin{equation}\label{gaugeres}
  \mathcal{G}_r \equiv \{g : \mathbb{R}^3\rightarrow G \mid g({\bf
  r})\in H \;\;\mbox{when} \mid{\bf r}\mid \rightarrow \infty\}.
\end{equation}
It is now important to distinguish the subgroup of
\textit{asymptotically trivial gauge transformations}
$\mathcal{H}_\infty $ defined as the group of residual
transformations which tend to the identity element at spatial
infinity:
\begin{equation}\label{gaugeinf}
  \mathcal{G}_\infty \equiv \{g : \mathbb{R}^3\rightarrow G \mid g({\bf
  r}) \rightarrow 1 \;\;\mbox{when} \mid{\bf r}\mid \rightarrow \infty\}.
\end{equation}
Clearly we may consider these transformations as maps from $S^3$
to $G$ with the "point at infinity" mapped to the identity element
of $G$.

Finally there is the \textit{connected component}
$\mathcal{G}_\infty^0 \subset \mathcal{G}_\infty $ of all elements
which can be continuously deformed to the identity element $g({\bf
r})\equiv 1$ of $\mathcal{G}_\infty $. The group
$\mathcal{G}_\infty^0 $ is a normal subgroup of
$\mathcal{G}_\infty $ consisting of the topologically trivial
gauge transformations. In general on has that
\begin{equation}\label{gaugequotient}
  \mathcal{G}_\infty /\mathcal{G}_\infty^0 \simeq \pi_3(G)
\end{equation}
saying that the elements of the coset are in one to one
correspondence with the homotopy classes of maps $S^3\rightarrow
G$ with $\infty\mapsto 1$. For abelian $G$ it follows that
$\mathcal{G}_\infty = \mathcal{G}_\infty^0 $, but for simple Lie
groups we have the result that $\pi_3(G) = \mathbb{Z}$. and there
are nontrivial residual gauge transformations. As will become
clear, these topologically nontrivial gauge transformations will
be related to the $\theta$ parameter which labels the nontrivial
ground states in non-abelian gauge theories.

Imposing Gauss' law on physical states in a given magnetic charge
sector means that we require the physical states in that sector to
be invariant under $\mathcal{G}_\infty^0 $. The result of the
analysis is that there is a group of physical, internal symmetry
transformations that generates physically inequivalent
configurations and therefore should lead to physical zero modes
and will upon quantization be realized on the physical states. It
is the group of residual gauge transformations modulo the
transformations generated by Gauss' constraint
\begin{equation}\label{gaugeinternal}
 \mathcal{S} \equiv \mathcal{G}_r /\mathcal{G}_\infty^0.
\end{equation}

A precise analysis of Balachandran and Giulini
\cite{Balachandran:1992rb,Giulini:1995bi} led to the following
structure for the situation for the 't Hooft-Polyakov monopoles in
the Georgi-Glashow model.
\begin{equation}\label{internal}
  \mathcal{S} = \left\{\begin{array}{cc}  \mathbb{Z} \times U(1) & n=0 \\
  \mathbb{Z}_{|n|} \times \mathbb{R} & n \neq 0
  \end{array} \right.
\end{equation}
 In the trivial sector this leads to two physical parameters
 referring to the representation labels of $\mathcal{S}$, but
 because $\mathcal{S}$ is defined as a quotient, the parameterization
 of the transformations is not trivial. It should take
 care of the way the $\theta$ parameter enters in the charged sectors
 of magnetic monopoles, the socalled Witten effect.


\subsection{The Witten effect: CP violation in the monopole sector}

The Witten effect\cite{Witten:1979ey} refers to the shift of the
allowed electric charges carried by magnetic monopoles. In other
words, the dyonic spectrum of the theory depends on the CP
violating $\theta$ vacuum parameter introduced by 't Hooft. The
$\theta$ parameter enters through the addition of a topological
term
\begin{equation}\label{theta}
   \mathcal{L}_\theta = \frac{\theta e^2}{32\pi^2}\;\; F\wedge F
\end{equation}
to the Lagrangian. This term is a total derivative and therefore
does not affect the field equations. However, on the quantum level
it does affect the physics and leads to an additional physical
parameter in the
theory\cite{'tHooft:1976up,'tHooft:1976fv,Jackiw:1976pf,Callan:1976je}.

Witten considered the implementation of the nontrivial internal
U(1) transformations on the fields and calculated the Noether
charge associated with that symmetry; and found that there is a
contribution from the $\theta$ term in the Lagrangian to that
$U(1)$ current.

\begin{figure}[!htb]
\begin{center}
\makebox[7.0cm]{\psfig{figure=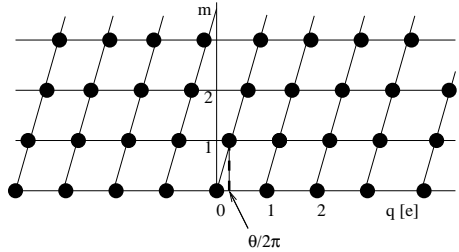,width=6.cm,angle=0}}
\caption[Hedgehog]{\footnotesize The figure shows the shift of
  electric charges in the magnetic sectors, due to the CP violating
  $\theta$ angle.} \label{thetalat.eps}
\end{center}
\end{figure}

In view of the observations made in the previous subsection we
focus on nontrivial gauge transformations which are constant
rotations generated by the Higgs field (normalized at infinity)
\begin{equation}\label{higgstransform}
  g(\textbf{r}) = \exp{[\alpha \Phi(\textbf{r})/f]}\;\;\mbox{ with
  } \;\;g_\infty(\hat{\mathbf{r}}) =
  \exp{[\alpha \hat{\Phi}_\infty(\hat{\mathbf{r}})]}
\end{equation}
Applying this transformation gives the infinitesimal changes in
the fields
\begin{equation}\label{variation}
\left\{ \begin{array}{l}\mathbf{a} = -\frac{1}{ef}
\;\;\mathbf{D}\Phi \\
\phi = 0 \end{array} \right.
\end{equation}
Now we may calculate the corresponding  charge generator $N$ from
Noethers theorem:
\begin{equation}\label{noether}
  N =  \int d^3\mathbf{r} [ \frac{\partial\mathcal{L}}{\partial(\partial_0
  \mathbf{A})}\cdot \mathbf{a} + \frac{\partial\mathcal{L}}{\partial(\partial_0
  \Phi)}\phi]
\end{equation}
Substituting the gauge transformations one obtains
\begin{equation}\label{charge}
  N=\int d^3\mathbf{r} [(\mathbf{E} -\frac{\theta
  e^2}{8\pi^2}\;\;\mathbf{B})\cdot \mathbf{D}\Phi ] =
  \frac{1}{e}(q
  - \frac{\theta e^2}{8\pi^2}\;\; g)
\end{equation}
this operator has an integer spectrum, from which the allowed charges
in the topological sector $m$ follow
\begin{equation}\label{chargeshift}
  q = (n + \frac{\theta m}{2\pi})\;\;e
\end{equation}
The electric charges in the dyonic sectors are shifted in a way
consistent with the $2\pi$ periodicity of $\theta$. The shift does
not violate the quantization condition for dyons, $q_1g_2-q_2g_1=
2\pi n$, because the $\theta$ dependent terms cancel. We have
indicated the $\theta$-dependent shift in the electric-magnetic
charge lattice (for SU(2))in Fig.\ref{thetalat.eps}.

An interesting consequence of this shift in the electric charge
proportional to theta and the magnetic charge, is the effect of
\textit{oblique confinement}\cite{}. If on considers the vacuum of
the unbroken gauge theory as a magnetic superconductor in which
the 't Hooft-Polyakov monopoles are condensed, one may wonder what
would happen to the condensate if one increases theta. Clearly the
total Coulombic repulsion of the monopoles in the vacuum will
increase, and moreover, if theta reaches $\pi$ the dyon with
smallest negative charge will have a smaller total charge and
therefore allow for a new ground state with lower energy. In other
words, this reasoning suggests a phase transition for $\theta
=\pi$ and shows what the physical nature of the transition is. In
principle one could imagine also other points on the lattice to
condense leading to a variety of \textit{oblique} confining
phases\cite{'tHooft:1981ht}.


\subsection{Isodoublet modes: Spin from Isospin}

So far we have looked at the Georgi-Glashow model without introducing
additional matter multiplets. It is interesting to do so and to
consider how an additional scalar or spinor doublet in the theory
affects the moduli space.

We have alluded before to the the spherically symmetric solutions.
These where $SO(3)$ symmetric with respect to a generator
$\mathbf{J}=\mathbf{L}+\mathbf{T}$ that simultaneously generates
space rotations and rigid gauge transformations. In studying the
modes for matter fields that couple to the monopole one finds that
it is $\mathbf{J}$ that generates the "true" angular momentum,
i.e. it is the operator that commutes with the Hamiltonian of the
system in the monopole background. This is reminiscent of the
situation in  abelian monopole physics where the
electromagnetic field of a dyonic system of a pole of strength $g$
and a charge $q$ carries an angular momentum $qg/4\pi$
(independent of the separation) in a direction pointing from the
charge to the pole. One finds that for a charge-pole pair
satisfying the minimal Dirac condition $qg=2\pi$ the possibility
of half integer "intrinsic" angular momentum arises. This
possibility is also naturally present in the model we have been
discussing where the 't Hooft-Polyakov monopole was doubly
charged, i.e. $eg=4\pi$, this is indeed consistent with the
presence of doublet fields which have electric charges $q=\pm e/2$
yielding the minimal value $qg=2\pi$. Detailed calculations from
't Hooft and Hasenfratz\cite{Hasenfratz:1976gr} and  Jackiw and
Rebbi\cite{Jackiw:1976xx} showed that for the scalar doublet modes
the generator of the angular momentum does indeed  have half
integral eigenvalues. Adding to the Lagrangian (\ref{action}) the
term for a doublet $U$
\begin{equation}\label{scalardoublet}
   \mathcal{L}_d = |D_\mu U|^2
\end{equation}
with $D_\mu U = (\partial_\mu +ieA_\mu^a \tau^a/2)U$. The
classical doublet mode can be written like
\begin{equation}\label{doubletmode}
  U= u(r) \exp( -i\alpha_a \tau^a/2) \mathbf{s}
\end{equation}
with three parameters arbitrary parameters $\alpha^a$and
$\mathbf{s}$ some constant spinor . The contribution of
$\mathcal{L}_d$ to the angular momentum generator is given by
\begin{equation}\label{angularmomentum}
\mathbf{J}= -\int\,d^3r\; \mathbf{r}\times [\Pi_U^\dagger(\nabla +
ie\mathbf{A}^a\tau^a/2)U + h.c.]
\end{equation}
where the conjugate momentum to U is $ \Pi_U = D_0U$. Making the
collective coordinates $\alpha$ in the solution time dependent,
i.e. $\alpha = \alpha (t)$ one obtains that $ \Pi_U^\dagger =
D_0U^\dagger = \dot{U}^\dagger$ yielding
\begin{equation}\label{angular2}
\mathbf{J}= -\int\,d^3r\; \dot{U}^\dagger(-\mathbf{r}\times\nabla
+ i\mathbf{\tau}/2)U + h.c.
\end{equation}
where we have used the background monopole solution. With the
spherical symmetry of the mode expression (\ref{doubletmode}) for
$U$ one is left with the last term in the bracket. So the physical
angular momentum $\mathbf{J}$ becomes equal to $\mathbf{T} =
\mathbf{\tau/2}$, the generator of the internal transformation,
which for the doublet is a (iso)spin one-half representation, i.e.
$s=qg/4\pi=1/2$ . And, as advertised, isospin has turned into
spin.


\subsection{Fermions from bosons}

If we succeeded in converting half-integral isospin to half-integral
spin we should also ask whether we have at the same time constructed
fermions out of bosons. In other words, we have to consider the
interchange properties of two dyonic composites. Let us rephrase an
ingenious argument originally due to Goldhaber\cite{Goldhaber:1976dp}.
The argument runs as follows. One considers two dyonic composites with
electric charge $e$ and magnetic charge $g$ with coordinates
$\mathbf{r_1}$ and $\mathbf{r_2}$.  The corresponding two particle
Schr\"odinger problem, can be separated in a center of mass and a
relative coordinate $ \mathbf{r} =\mathbf{r_1}-\mathbf{r_2}$. The part
of the wave function depending  on the relative coordinate is of course
defined on the space of this relative coordinate. But as the particles
are considered to be indistinguishable we may identify the points
$\mathbf{r}$ and $\mathbf{-r}$. We furthermore keep the dyons well
separated so that the interiors do not take part in the interchange,
which means that we exclude the point $\mathbf{r}=0$. The resulting two
particle space (taking $r$ fixed) is then topologically equivalent to a
two-sphere with opposite points identified i.e. two-dimensional real
projective space. A physical interchange corresponds to a closed path
in this projective space, and since the first homotopy group
$\pi_1(P\mathbb{R}_2) = \mathbb{Z}_2$ there are two inequivalent
classes of wave functions on this space which correspond to the
different representations of this $\mathbb{Z}_2$. This is the
topological origin of the quantum mechanical exchange properties of
particles. If there were no electromagnetic field present, the
arguments allows one to introduce fermions and bosons as the particles
to start of with. In our case these are chosen to be bosons. The
electromagnetic interaction term with charge monopole interactions
included will take the following form
\begin{equation}\label{relative}
  \mathbf{p} +
  ie[\mathbf{A}(\mathbf{r})-\mathbf{A}(\mathbf{-r})]\;.
\end{equation}
The effective gauge potential in brackets has zero magnetic field
because one subtracts the field in the point $\mathbf{r}$ with the
field in the point $\mathbf{-r}$ and these are equal up to a
(topologically nontrivial) gauge transformation\footnote{Take for
example the potentials as in Eqn.{\ref{diracpotential}}, then it is
clear that $\mathbf{a}_{\hat{n}}(-\mathbf{r}) =
\mathbf{a}_{-\hat{n}}(\mathbf{r})$ }. Indeed the overall
electromagnetic interaction between two particles that carry electric
and magnetic charges with the same ratio, are strictly dual to two
purely electric charges which have only Coulombic interactions. So
there is no net magnetic field in the configuration and the total
expression between the brackets must indeed be pure gauge. This
implies that by taking the relative particle coordinate around a closed
loop (i.e. interchanging the two particles) only two things may happen
to the phase of the wave function; because the  phase cannot dependent
on continuous deformations of the path taken, it can only depend on the
homotopy class of the path. In particular since the space is doubly
connected the resulting gauge connection may only induce a nontrivial
$\mathbb{Z}_2$ phase if one goes to the opposite point on the sphere,
i.e takes a non-contractible loop. This is indeed the case. In the
original non-trivial $U(1)$ monopole bundle one needs two overlapping
patches, say northern and southern hemispheres, The gauge potentials in
the different patches differ by a gauge transformation
\begin{equation}\label{patches}
  e\mathbf{A}_{II} = e\mathbf{A}_{I} + \mathbf{\nabla} \chi
\end{equation}
with the single valuedness condition in the overlap yielding
$\chi(\varphi+2\pi) - \chi(\varphi) = 2\pi m = eg$. The two-particle
wave functions are  sections of a (non)trivial $\mathbb{Z}_2$ bundle
over $P\mathbb{R}_2$ with transition function $\exp (ieg/2)=(-1)^{2s}$.
The spin-statistics connection is saved from a painful demise, and we
succeeded in making fermions out of bosons.

\subsection{The Rubakov-Callan effect:\\monopole induced catalysis of
baryon number violation}

If a charge moves straight towards a monopole no force is exerted
on it and therefore it would just move straight through the
monopole. But if this is what happens, the total angular momentum
$J$ would not be conserved  because in this process $L=0$ and the
other term $eg\hat{r}/4\pi$ would change sign. This contradiction
is resolved in more realistic quantum mechanical descriptions. For
example if we consider a Dirac field in the abelian, minimal
($eg/4\pi = 1/2$) monopole background we find that in the lowest
angular momentum state the radial component of the spin has to
satisfy\cite{Kazama:1977fm},
\begin{equation}\label{radialspin}
  \sigma\cdot\hat{\mathbf{r}} = \frac{eg}{|eg|}\;,
\end{equation}
and hence the helicity operator $h$ is related to the charge and
equals
\begin{equation}\label{helicity}
  h= -\frac{eg}{|eg|}\hat{p}_r \;.
\end{equation}
This implies that scattering of a fermion by a monopole in the
abelian theory always induces a helicity flip. Though naively the
helicity operator commutes with the Hamiltonian and therefore
should be conserved, the fact is that careful analysis of the
self-adjointness property of the Hamiltonian, leads to boundary
conditions for which the helicity operator is not hermitean and
therefore helicity needs no longer to be
conserved\cite{Goldhaber:1977xw}.

In the case of the 't Hooft-Polyakov monopole however, the
situation turns out to be radically different. Now one has to
return to the analysis of a spinor iso-doublet field in the
monopole background. The scattering
solutions\cite{Marciano:1983md} in the $J=0$ channel describe a
process where charge exchange is the mechanism by which the
angular momentum conservation is saved. The cross-section is
basically determined by kinematics and will be of order $\pi/k^2$.
One may employ the same analysis to the fundamental $SU(5)$
monopole to find out that baryon number would not be conserved in
such a process. The minimal monopole typically couples to doublets
$(d_3,e^+)$ and $(u_1^c,u_2)$ (where the numerical indices refer
to color), so in general one obtains processes like:
\begin{eqnarray}\label{bviolation}
d_{3R} + dyon &\rightarrow &  e_R^+ + dyon \nonumber \\
e^+_{R} + dyon &\rightarrow &  d_{3l} + dyon \nonumber \\
u^c_{1R} + dyon &\rightarrow &  u_{2R} + dyon \\
u_{2L} + dyon &\rightarrow &  u^c_{1L} + dyon \;.\nonumber
\end{eqnarray}
In this mode analysis also electric charge and color conservation
appear to be violated, which is of course an artifact because the
back reaction of the monopole is not taken care of. In a full
quantum mechanical treatment one finds that the charge degrees of
freedom on the monopole will be excited as to ensure conservation
of the charges related to local symmetries. But the violation of
baryon and lepton number remains. Rubakov and Callan took the
previous analysis some dramatic steps further. The studied the
full quantum dynamics in the spherically symmetric $J=0$ channel,
in which the fermion monopole system reduces to a two-dimensional
chiral Schwinger model on a halfline. This model has been analyzed
in detail in the fermionic formulation by Rubakov
\cite{Rubakov:1981rg,Rubakov:1982fp} and by Callan
\cite{Callan:1982au,Callan:1982ah} after bosonisation. Most
remarkable is the persistence of the large geometric cross section
which is not cutoff by the symmetry breaking scale, therefore
baryons can decay with dramatic rates at low energy in the
presence of grand unified monopoles.


\section{Supersymmetric monopoles}

The study of monopoles has become a crucial ingredient in understanding
the physical properties of non-abelian gauge theories.  Remarkable
progress has been made in particular in the realm of suppersymmetric
gauge theories where a number of important exact results have been
obtained. In the following subsections we review some of the turning
points along these lines of development. To fix the setting, let us
consider the $N=2$ supersymmetric $SU(2)$ Yang Mills theory with
Lagrangian
\begin{equation}\label{susy}
  \mathcal{L} = -\frac{1}{4}F^2 + \bar{\Psi}D\!\!\!\!/\Psi +
  \frac{1}{2}(D\phi)^2+\frac{1}{2}(D \chi )^2 -e\bar{\Psi}[\phi +
  i\gamma_5\chi ,\Psi] + \frac{1}{2} e^2 [\phi,\chi ]^2
\end{equation}
The theory involves one N=2 chiral (or vector) supermultiplet in
the adjoint representation of the gauge group.  This multiplet
consists of a the gauge field $A_\mu$, two Weyl spinors combined
in a single Dirac spinor $\Psi$, a scalar $\phi$ and pseudoscalar
$\chi$.  Alternatively one may choose to keep the two Weyl spinors
and combine both scalar components in a single complex scalar
field. In this formulation the potential is just given by
$\frac{1}{2} e^2[\Phi,\Phi^\dagger]^2$. Setting the fields $\Psi$
and $\chi$ to zero we recover the bosonic Georgi-Glashow model in
the BPS-limit, with the well known monopole solution. At this
point it is convenient to rescale the whole supermultiplet by the
coupling constant $e$, giving an overall factor $1/e^2$ in front
of the Lagrangian.

It is particularly interesting to study the fermionic zero modes
that have to satisfy
\begin{equation}\label{zeromodes}
  i D\!\!\!\!/ \Psi -[\phi,\Psi] = 0 \; .
\end{equation}
Solutions can be generated applying a supersymmetry transformation
to the classical, bosonic monopole configuration and read of what
it yields for the spinor. The transformation rule gives
\begin{equation}\label{spinortransform}
  \delta\Psi =(\sigma_{\mu\nu}F^{\mu\nu} - D\!\!\!\!/\phi +\gamma_5[\chi,\phi]
  +i\gamma_5 D\!\!\!\!/\chi )\varepsilon
\end{equation}
Using the Bogomol'nyi equations only one component of the first
two terms survives:
\begin{equation}\label{spinor}
\Psi =\left( \begin{array}{c}
  \chi^+ \\
  \chi^-
\end{array}\right) = \left( \begin{array}{cc}
  \mathbf{\sigma}\cdot\mathbf{B} & 0 \\
  0 & 0
\end{array}\right) \left(\begin{array}{c}
  s^+ \\
  s^-
\end{array}\right)
\end{equation}
and one obtains  the solution as first discussed by Jackiw and
Rebbi,
\begin{equation}\label{chispinor}
\chi^+ = \mathbf{\sigma}\!\cdot\mathbf{B}\;s^+
\;\;\;\mbox{and}\;\;\; \chi^- = 0
\end{equation}
Here the two-component spinor $s^+$ is still arbitrary so that we
end up with two zeromodes,  which are furthermore charge
conjugation invariant. These modes of course also exist in
non-supersymmetric versions of the model. In the expansion of the
quantized Dirac field these modes have to be included, we have to
write
\begin{equation}\label{psi}
  \Psi = c_1 \psi_1 + c_2\psi_2 + \sum_p(b_p\psi_p + d_p^\dagger
  \psi_p^c).
\end{equation}
The anticommutator of $\Psi$ and $\Psi^\dagger$ imposes that the
$c$ operators obey a Clifford algebra
$\{c_i,c^\dagger_j\}=\delta_{ij}$. This algebra has a
4-dimensional representation with states that have the
properties indicated in the following table:\\[2mm]
\begin{center}
\begin{tabular}{|l|c|c|c|c|}
  \hline
 State & $\mid ++>$ & $\mid +->$ & $\mid -+>$ & $\mid -->$ \\
 \hline
  Fermion number& 1 & 0 & 0 & -1 \\
  \hline
  Spin & 0 & \multicolumn{2}{|c|}{$\frac{1}{2}$} & 0 \\
  \hline
\end{tabular}\\[2mm]
{\it The 4-fold ground state degeneracy of the N=2 supersymmetric
monopole}
\end{center}
\vspace*{5mm} The conclusion is thus that the groundstate of the
N=2 supersymmetric monopole is 4-fold degenerate.


\subsection{ The supersymmetry algebra with central charge}

Witten and Olive\cite{Witten:1978mh}studied the way monopoles
behave in supersymmetric gauge theories from a different angle. An
immediate motivation is the fact that the BPS limit is implied by
supersymmetry. By studying the super-algebra and its
representations they obtained the remarkable result that states
saturating the Bogomol'nyi bound remain doing so on the quantum
level. The reason is that the BPS states have to form special
so-called \textit{short} representations of the algebra.

The N=2 algebra of super charges allows for central charges $U$
and $V$ and has the following general form:
\begin{equation}\label{superalgebra}
  \{Q_{\alpha i}, \bar{Q}_{\beta j}\} = \delta_{ij}
  \gamma^mu_{\alpha\beta}P_\mu + \varepsilon_{ij}(\delta_{\alpha\beta} U
  +(\gamma_5)_{\alpha\beta}V)
\end{equation}
The two charges $Q_\alpha$ are two-component Majorana spinors. In
the monopole state one obtains that the central charge $V$ is
nonvanishing
\begin{equation}\label{centralcharge}
  V = \int\; d^3 x [\;\phi(\nabla\!\cdot\mathbf{B}) +
  \chi(\nabla\!\cdot\mathbf{E})]= gf \geq 0
\end{equation}
$U$ is obtained by interchanging $\Phi$ and $\chi$ and equals zero
in the pure monopole case. Evaluating the algebra in the rest
frame gives
\begin{equation}\label{restframe}
\{Q_{\alpha i}, \bar{Q}_{\beta j}\} = \delta_{ij}
  \delta_{\alpha\beta}M + \varepsilon_{ij}(\gamma_5\gamma_0)_{\alpha\beta}\;\;gf
\end{equation}
Positivity of the anticommutator leads to the bound $M^2\geq
(gf)^2$. Something special happens if the bound is saturated, then
the representation theory of the N=2 algebra alters and allows for
a so called {\it short} representation of $2^N = 4$ states, with
spin content $(\frac{1}{2},0^+,0^-)$. Normal, "\textit{long}"
representations have $2^{2N}$ states. We see that the zeromodes
discussed in the previous section indeed form a short
representation. And what about possible excited dyonic bound
states one might wonder. These do exist and have been
analyzed\cite{Bais:1981gv} and give - as expected - rise to
\textit{long} representations because they do not correspond to
zeromodes and hence do not satisfy the Bogol'nyi bound. This
distinction between short and long representations can be compared
to the difference between massive and massless representations of
the Lorentz group. In the quantum BPS limit of the N=2 theory we
arrive at the conclusion that the monopole groundstate forms a
scalar multiplet (containing a spin 1/2 doublet). Osborn extended
this analysis to the $N=4$ case obtaining that the lowest monopole
states form the short N=4 multiplet now containing 16 states
consisting of a vector, four spinors and six scalars, i.e. a
magnetic copy of the vector multiplets that define the
theory\cite{Osborn:1979tq}.


\subsection{Duality regained}

In 1979 Montonen and Olive\cite{Montonen:1977sn}made a daring
conjecture concerning spontaneously broken gauge theories.
Inspired by the the mass formula for BPS states they conjectured
that --- in the full quantum theory --- there would be a
non-abelian version of electric magnetic duality realized in the
supersymmetric $SO(3)$ model. This duality would be a four
dimensional analogue of the equivalence of the Sine--Gordon Theory
and the massive Thirring model in two dimensions. The strongest
form of it would be realized in the N=4 theory. In this case the
quantum numbers of the electric and magnetic BPS states are such
that the model could be \textit{self}-dual in the sense that the
massive vector bosons would be mapped on the monopoles and vice
versa, implying that the charges get also mapped onto each other,
i.e. $e \leftrightarrow g= 4\pi/e$. This is clearly a strong--weak
coupling duality in that the fine structure constant would be
mapped to its inverse. The conjecture amounts to the statement
that in the strong coupling regime the physics would be described
by a weakly interacting spontaneously broken "magnetic" gauge
theory with a dual gauge group $\tilde{G}\simeq SO(3)$. There
appears considerable evidence for this idea:
\begin{description}
  \item[-]The BPS mass relation $m = f \sqrt{q^2 + g^2}$.
  \item[-]The BPS mass formula receives no quantum corrections as
  it appears in the central charge of the supersymmetry algebra.
  \item[-]In the model there is no charge renormalization.
  \item[-]The spin content of the electric and magnetic representations is the
    same (1 vector, 4 spinors and 6 (pseudo)scalars).
\end{description}

This bold idea has in recent years developed into a broad research
field mainly due to the work of Seiberg and Witten on
electric--magnetic dualities in general supersymmetric gauge
theories. In the following sections we briefly summarize some
aspects of this work.

\subsection{$\mathbf{SL(2,\mathbb{Z})}$ duality}

We may extend the Montonen--Olive duality  to a full
$SL(2,\mathbb{Z})$ duality by including also the $\theta$
parameter. Let us introduce the complex coupling parameter
\begin{equation}\label{tau}
  \tau\equiv\frac{\theta}{2\pi} + \imath\frac{4\pi}{e^2}
\end{equation}
We recall formula that for the dyons with magnetic charge $g=4\pi
m/e$ in the presence of the $\theta$ parameter the electric
charges are given by $q = ne + m\theta e/2\pi$. The exact BPS mass
formula as it follows from the central charge of supersymmetry
algebra can now be casted in the form
\begin{equation}\label{massformula}
  M(m,n) = ef|n+m\tau| \;\;\; m,n\in \mathbb{Z}
\end{equation}
This BPS mass formula is invariant under  $SL(2,\mathbb{Z})$
transformations which act as follows. Let $M$ be an element of
$SL(2,\mathbb{Z})$, which is defined as
\begin{equation}\label{sl2z}
  M=\left(\begin{array}{cc}
    a & b \\
    c & d \
  \end{array}\right), \;\;\;\mbox{with}\; a,b,c,d \in \mathbb{Z}
  \;\; \mbox{and} \; ad-bc=1 ,
\end{equation}
then the parameter $\tau$ transforms as
\begin{equation}\label{tautransform}
  \tau\rightarrow\tau\prime =\frac{a\tau +b}{c\tau +d},
\end{equation}
whereas the BPS states transform as
\begin{equation}\label{bpstransform}
  (m,n)\rightarrow (m',n') = (m,n)M^{-1} = (m,n)\left(\begin{array}{cc}
    d & -b \\
    -c & a \
  \end{array}\right).
\end{equation}
 The group $SL(2,\mathbb{Z})$ is generated by $T:
\tau\rightarrow\tau+1$ and $S = \tau\rightarrow -1/\tau$. The
generator $T$ generates a shift of $\theta$ over $2\pi$, while for
$\theta=0$ the $S$ generator corresponds to the original
Montonen--Olive duality transformation, with $e\rightarrow 4\pi/e$
and $(m,n)\rightarrow (-n,m)$. The $SL(2,\mathbb{Z})$-duality
conjecture reads that a theory would be invariant under the
$SL(2,\mathbb{Z})$ transformations  on the dyonic spectrum of BPS
states.

From the mass formula one may deduce a stability condition. One
easily sees that the following inequality between different mass
states holds
\begin{equation}\label{decay}
  M(m_1+m_2,n_1+n_2) \leq M(m_1,n_1) + M(m_2,n_2)
\end{equation}
implying that the strict equality holds if and only if $m_1+m_2$
and $n_1+n_2$ are relative prime. A dyon with magnetic and
electric quantum numbers which are relative prime should therefore
be stable, though as we will see later on, the stability depends on which
extended supersymmetry is realized.


\subsection{The Seiberg--Witten results: the N=2 vacuum structure}

The question what  happens to electric--magnetic duality for N=2
or N=1 supersymmetric gauge theories was answered by some
remarkable exact results obtained by Seiberg and Witten
\cite{Seiberg:1994rs,Seiberg:1994aj}. The N=2 theory is of course
asymptotically free, allowing a choice of scale. The question is
to understand the vacuum structure of the theory and the
dependence of the BPS spectrum on the vacuum parameters. The
classical potential is given by
\begin{equation}\label{susypotential}
  V(\Phi) = \frac{1}{e^2}[\Phi,\Phi^\dagger]^2\;,
\end{equation}
consequently there is a continuous family of classical vacua
because for the potential to acquire its minimum it suffices for
$\Phi$ and $\Phi^\dagger$ to commute. There is a global $U(1)$ R
symmetry of the classical potential on the classical level which
gets broken by instanton effects to a global $\mathbb{Z}_8$. The
scalar field is doubly charged under this discrete group. A
constant value $\Phi = a\tau_3/2$ breaks the gauge group to a
$U(1)$.  A gauge invariant parameterization of the classical vacua
is  given by the complex parameter
\begin{equation}\label{uparameter}
  u = \frac{1}{2}a^2 = Tr\Phi^2
\end{equation}
Because $u$ has charge $k=4$ under the $\mathbb{Z}_8$ group, this
symmetry gets broken to a global $\mathbb{Z}_2$ which acts on  the
moduli space as $u\rightarrow -u$. On a classical level there is
only one special point, the point $u=0$ where the full $SU(2)$
symmetry is restored.

The quantum moduli space parameterized by the the vacuum
expectation value $u=<Tr\Phi^2>$ depends on the effective action,
which in turn depends on a single holomorphic function
$\mathcal{F}(a)$. Witten and Seiberg succeeded in explicitly
calculating $a(u)$ and $a_D\equiv (\partial\mathcal{F}/\partial
a)(u)$ in terms of certain hypergeometric functions from the
holomorphicity constraint combined with certain information on the
possible singularities and some (one loop) perturbative results.
This information is crucial because it determines the masses of
the BPS states as a function of $u$ again from the central charge
in the N=2 supersymmetry algebra as
\begin{equation}\label{quantumbps}
  M(m,n) = \sqrt{2}|na(u)-ma_D(u)|
\end{equation}
We see that classically $\tau_{cl} = a_D/a$, however on the
quantum level $\tau$ is rather defined as $\tau
=(\partial^2\mathcal{F}/\partial a^2)(u)$. The parameters $a$ and
$a_D$ are both complex and generically the masses can be
represented on a lattice where the mass just corresponds to the
Euclidean length of the corresponding lattice vector $(m,n)$. The
situation is similar to the one we described before where the
triangle inequality determines the stability of various states in
the lattice. The interpretation of the phase structure and the BPS
spectrum depends critically on the singularities of $a$ and $a_D$
in the $u$ plane. There are branchpoints for $u=\pm 1$ and
$u=\infty$ with cuts along the (negative) real axis for $a_D$
extending to $u=-1$ and for $a$ to $u=1$. We see that the
singularities have moved away from the classical point $u=0$. From
the singularity structure one may obtain the monodromies generated
on the vector $(a,a_D)$ by going around the singularities. The
requirement that the physical spectrum should be invariant under
the action of the monodromy group will put constraints on which
states are allowed.

\begin{figure}[!htb]
\begin{center}
\makebox[10.0cm]{\psfig{figure=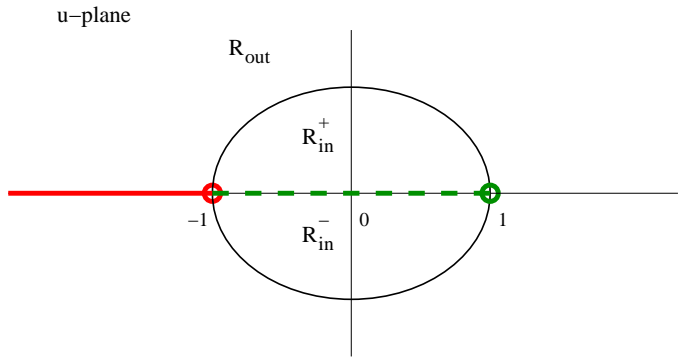,width=9.cm,angle=0}}
\caption[Hedgehog]{\footnotesize This figure shows the singularity
  structure of $a(u)$ and $a_D(u)$. On the curve the quotient $a_D/a$ becomes
  real, massless states only occur at $u=\pm 1.$}
\label{uplane.eps}
\end{center}
\end{figure}

Of particular interest are the points in the $u$ plane where
$a_D/a \in \mathbb{R}$ because then the lattice degenerates and
instabilities as well as massless states may develop. This
situation arises for a locus of points which form a closed curve
$\mathcal{C}$, called the curve of marginal stability. We have
depicted the situation in Fig.\ref{uplane.eps}. On $\mathcal{C}$
the ratio $a_D/a$ takes on all values in the interval $[-1,1]$.
The large $u$ region is the weakly coupled semiclassical region
whereas in the interior domain near and inside the curve of
marginal stability, we are dealing with the strong coupling
regime. It is clear that we have to distinguish three different
regions, the region outside the curve $R_{out}$, and inside the
curve the regions above and below the real axis, $R_{in}^\pm$.
The interpretation of the curve is as follows. Clearly as long as
we do not cross it the spectrum will depend smoothly on $u$ and
cannot develop any instabilities, neither inside, nor outside the
curve. It follows from the mass formula (\ref{quantumbps}) that on
the curve massless states may develop. For $M=0$ we see that we
must have $n/m=a_D/a \in [-1,1]$. In fact naively, that means
semi-classically, one would expect that any state with $n/m \in
[-1,1]$ will become massless exactly at the corresponding point on
$\mathcal{C}$ where $a_D/a(u)=n/m$.  One clearly expects that the
effective action and therefore $a$ and $a_D$ would develop
singularities, but this --- as followed from the result of Seiberg
and Witten --- only happens for $u=\pm 1$. Conversely if a
massless state would develop for some point $u$ in the moduli
space it would give rise to a singularity but there are no other
singularities than the ones just mentioned, which leads us to conclude that such states should be
excluded. The point u=+1 corresponds to a massless state $(0,\pm
1)$ a monopole or anti-monopole, whereas the point $u=-1$
corresponds to the (anti-)dyons $\pm(\pm 1,1)$. These points
correspond to groundstates where monopoles/dyons become massless
and hence will condense to realize the confining phases
anticipated by Mandelstam and 't Hooft in the seventies. A
striking result. Adding to this the monodromy constraints
mentioned before, the result for the $N=2$ theory without extra
matter multiplets can be summarized as follows. In the weak
coupling regime the spectrum consists only of the states $\{\pm
(1,0), \pm (n,1); n\in \mathbb{Z}\}$. In the strong regime similar
arguments\cite{Ferrari:1996sv,Bilal:1996sk} yield that the
spectrum is further reduced to just states that become massless
$\{\pm (01), \pm(-1,1)\}$ for the region above and $\{\pm (01),
\pm(-1,1)\}$ for the region below the real axis.
\section{Beyond $SU(2)$: gauge groups and integrable systems}
The construction of general multi-monopole monopole solutions to
the Bogomol'nyi equations for a theory with some general gauge
group has been a source of inspiration for  a full generation of
mathematical physicists. Substantial progress has been made in
constructing explicit solutions, and many intriguing relations
have been found with various classes of integrable systems in
lower dimensions. From these analogies, various solution
techniques for (multi)monopoles have been developed especially if
additional symmetries were imposed, lowering the effective
spacetime dimensionality of the system. This in a first wave gave
rise to obtaining multi-monopole solutions for gauge group SU(2),
for example by exploiting Bäcklund transformation
techniques\cite{Forgacs:1980ym,Forgacs:1981ve}. An general
framework to discuss the selfduality and Bogomol'nyi equations
based on algebraic geometry and the use of twistors was introduced
by Ward\cite{Ward:1977ta,Ward:1981jb} and Atiyah and
Ward\cite{Atiyah:1977pw}. Another general approach, based on the
theory complex vector bundles, developed to solve the general
instanton problem, usually referred to as the
Atiyah-Hitchin-Drinfeld-Manin construction\cite{Atiyah:1978ri},
was extended by Nahm for the general Bogomol'nyi equations (see
below).

\subsection{Spherical symmetry: Toda systems}
The simplest class of monopoles for which the exact solutions were
constructed is the class of {\it spherically symmetric} solutions,
where one defines a generalized $SO(3)$ generator
$\mathbf{J}=\mathbf{L}+\mathbf{T}$ and imposes the
conditions\ref{spherical} on the fields. In the present context
$\mathbf{T}$ generates some $SU(2)$ subgroup of the the full gauge
group. If one chooses the minimal regular embedding corresponding
to the (simple) roots of the full gauge algebra one obtains the
fundamental monopoles discussed before. The choice of a
non-regular embedding leads to more interesting results. For
example, the choice of the {\it maximal} or better {\it principal}
$SU(2)$ embedding, where the fundamental (vector) representation
branches to a single irreducible $SU(2)$ representation leads to
one- (two-) dimensional conformally invariant Toda-like
systems\cite{Bais:1978yh,Wilkinson:1979zh,Leznov:1979kh}. The
Bogomol'nyi system reduces to
\begin{equation}\label{toda}
\partial \bar{\partial} \;\rho_i = \sum_{j=1}^\ell K_{ij}\;
e^{\rho_j}\;,
\end{equation}
where we have introduced complex coordinates
\begin{equation}\label{coordinate}
  z= r + it  \;,\;\; \bar{z} = r- it.
\end{equation}
These finite Toda-like systems are completely characterized by the
Cartan matrix $K$ of the Lie algebra of $G$, defined as the
(asymmetrically) normalized inner product of the $\ell$ (= rank
$G$) simple roots $\gamma_a$
\begin{equation}\label{cartanmatrix}
  K_{ij} \equiv 2\;
{\overrightarrow{\gamma}_i \cdot \overrightarrow{\gamma}_j}/
  {(\overrightarrow{\gamma}_j \cdot \overrightarrow{\gamma}_j)}
\end{equation}
The $SU(2)$ version of this system is just the Liouville equation
initially obtained in this context by Witten\cite{Witten:1977ck}.
The complex equations yield the instanton solutions, the
restriction to $\rho_i = \rho_i(r)$ yields the monopoles. These
systems have returned in the context of conformal field theory
models for critical systems. The cases for arbitrary $SU(2)$
embeddings involve certain non-abelian generalizations of these
finite Toda systems\cite{Bais:1991bs}.

\subsection{Axial symmetry: nonlinear sigma models}
Let us briefly consider the axially symmetric case where one
imposes symmetry with respect to a single mixed space and gauge
rotation. The Bogomol'nyi equations now reduce to an integrable,
non-compact, non-linear sigma-model in a curved two-dimensional
$(\rho,z)$-space as followed from an analysis of Bais and Sasaki
\cite{Bais:1983av};
\begin{equation}\label{sigmamodel}
\nabla\left[ \rho(\nabla \mu) \; \mu^{-1} \right]= 0
\end{equation}
with $\nabla\equiv(\partial_\rho,\partial_z)$. The field variables
$\mu$ are defined as follows:
\begin{equation}\label{sigmafield}
  \mu\equiv g^\dagger g \;,\;\; g = g(\rho,z) \in G^* \;.
\end{equation}
The group $G^*$ is any non-compact real form  of $G$ based on a
symmetric decomposition of the Lie-algebra $G= K+P$, with
\[ [K,K]  \subset K \;\;,\;\; [P,K] \subset K \;\;,\;\; [P,P]\subset K \;.\]

As follows from its definition, $\mu$ is in fact an element of the
symmetric non-compact coset space $G^*/K$, where $K$ is the maximal
compact subgroup of $G^*$. For $SU(2)$ the system of Eqn.
(\ref{sigmamodel}) reduces to the case  studied by
Manton\cite{Manton:1978fr} and to the Ernst equation -- well known in
general relativity.
\subsection{The Nahm equations.}

We have mentioned in passing that classical Bogomol'nyi  or
self-dual Yang--Mills equations for monopole respectively
instantons differ primarily by their spacetime structure and
choice of boundary conditions. This means that the techniques to
solve them should be quite similar. The general instanton problem
was reduced to a entirely algebraic problem by Atiyah, Hitchin,
Drinfeld and Manin\cite{Atiyah:1978ri}, the so-called ADHM
construction. Nahm adapted this method  for the general Bogoml'nyi
equations, not surprisingly referred to as the ADHMN construction
\cite{Nahm:1981xg,Nahm:1981nb}. Nahm reduced the Bogomol'nyi
system to a system of of coupled nonlinear {\it ordinary}
differential equations for three $(n\times n)$ matrix valued
functions $T_i$ ($i=1,2,3$) of some variable $z$ defined on an
appropriate interval. The variable $z$ indicates the Fourier
components of the time variable $x_0$. The equations read:
\begin{equation}\label{nahm}
  \frac{d T_i}{dz} = i\epsilon_{ijk} [T_j,T_k]
\end{equation}
The way one arrives at this system of equations starts out with
the zero-modes $\psi^{(r)}(x,z)$ of the covariant equation for a
Weyl spinor in the fundamental representation of the gauge group
in the monopole background. To be definite we assume in the
following that the gauge group $G=SU(N)$\cite{Bowman:1983ss}. The
fundamental spinor has to satisfy:
\begin{equation}\label{spinoreqn}
D^\dagger\; \psi(\mathbf{x,z}) = 0
\end{equation}
with the operator defined as
\begin{equation}\label{D}
  D = \mathbf{\sigma}.(\nabla + ieA) - z + \Phi
\end{equation}
The matrices $T_i$ are defined as,
\begin{equation}\label{definitionX}
  T^{(rs)}_i(z) = \int\psi^{(r)\dagger}(x,z) \;x_i\;
\psi^{(s)}(x,z)\; d^3x
\end{equation}
The indices run $r$ and $s$ run over the number of zero-modes $n=
n(z)$ of the adjoint operator $D^\dagger$ in the given sector with
topological (magnetic) charge $m$. Clearly the number of
normalizable zero-modes depends on $z$, and consequently the
dimensionality of the $T$ matrices does as well. The tricky part
sits in the boudary conditions one has to impose on $z$, these
depend on the asymptotic behavior of the Higgs field,
\begin{equation}\label{asymptotics}
  \Phi \simeq \sum_{\ell=1}^N P_\ell\left[z_\ell +
  \frac{k_\ell}{2|\mathbf{x}|}\right]\;\; |\mathbf{x}|\rightarrow\infty
\end{equation}
The $P_\ell$ are spherical Legendre functions. In terms of the
parameters $z_\ell$ and $k_\ell$ the number of normalizable modes
to the spinor equation turns out to be
\begin{equation}\label{numberofmodes}
n(z) =\sum_{\ell=1}^N \; k_\ell \theta(z-z_\ell)\;.
\end{equation}
So $z$ ranges between the largest and smallest eigenvalues of the
Higgs field and $n(z)$ jumps at points where z becomes equal to an
eigenvalue of $\Phi$ (except if the eigenvalues happen to be
degenerate).

To reconstruct the original gauge and scalar fields given the
$T_i$ matrix functions, one goes in principle the opposite
direction. One first solves the linear equation:
\begin{equation}\label{inverse}
  \triangle^\dagger v_k(z,\mathbf{x})= 0
\end{equation}
with the $(n\times n)$ quaternionic operator given by
\begin{equation}\label{triangle}
 \triangle = i\partial_z + (\mathbf{x} + i\mathbf{T}) \cdot
 \mathbf{\sigma}\;.
\end{equation}
It can be shown that this equation has $N$ solutions for $SU(N)$.
Setting the normalized solutions $v_k$ in a $(n(z) \times N)$
matrix $V$ one may extract the  gauge potentials and scalar field
which solve the original Bogomol'nyi problem by computing
\begin{eqnarray}\label{inversenahm}
\mathbf{A} &=& \int dz \;V^\dagger \,\mathbf{\nabla} \, V \\
\Phi &=& \int dz \;zV^\dagger \, V
\end{eqnarray}
where we have normalized
\begin{equation}\label{nahmnorm}
  \int dz \;V^\dagger V = 1 \;.
\end{equation}
The Nahm equations form an integrable system and have popped up in
many places most recently in connection with D-brane descriptions
of monopoles in M-theory.

\section*{Conclusions and outlook}
\begin{quote}
{\it "Sag mir wo die Blumen sind, wo sind sie
geblieben..."\\\mbox{}\hfill M. Dietrich (1931)}
\end{quote}

We have reviewed the physics of magnetic monopoles in non-abelian
gauge theories. We have touched upon many different subjects, but
could not -- in the limited space available -- go into much
detail. Quite a few subjects were left out, in that sense the
particular choices made did not reflect the limited interest of
the author but rather his limited expertise. The study of
monopoles which received little encouragement from experiment
remains a highly interesting topic for theoretical research.
\textit{To be or not to be} remains for the time being, an open
question.

\textit{Acknowledgement}: The author wishes to thank many of his
(former) students and colleagues for inspiring collaborations on a wide variety of
questions concerning magnetic monopoles.

\bibliographystyle{unsrt}
\bibliography{monopoles}
\end{document}